\documentclass[showpacs,amsmath,amssymb,twocolumn,pra,superscriptaddress]{revtex4-2}

\usepackage{graphicx}
\usepackage{bm}
\usepackage{amsmath}
\usepackage[fleqn,tbtags]{mathtools}
\usepackage{graphicx}
\usepackage{physics}
\usepackage{amsfonts}
\usepackage[title]{appendix}
\usepackage{mathrsfs}
\usepackage[colorlinks=true, allcolors=blue]{hyperref}
\usepackage{tikz,pgfplots}
\usepackage{lipsum}
\usepackage{balance}

\usetikzlibrary{positioning}
\usetikzlibrary{decorations}
\definecolor{evergreen}{rgb}{0.27,0.62,0.20}

\tikzstyle{tensor_blue}=[rectangle,draw=black,fill=blue!25,thick,minimum size=0.5cm]
\tikzstyle{tensor_green}=[rectangle,draw=black,fill=green!20,thick,minimum size=0.5cm]
\tikzstyle{tensor_purple}=[rectangle,draw=black,fill=blue!50!red!50,thick,minimum size=0.5cm]

\newcommand{\RN}[1]{%
  \textup{\uppercase\expandafter{\romannumeral#1}}%
}

\definecolor{dullblue}{rgb}{.29,.47,.77}

\newtheorem{theorem}{Theorem}

\begin{document}

\preprint{APS/123-QED}

\title{Magic of Random Matrix Product States}

\author{Liyuan Chen}
\email{liyuanchen@fas.harvard.edu}
\affiliation{Department of Physics, Harvard University, Cambridge, Massachusetts 02138, USA}
\affiliation{John A. Paulson School of Engineering and Applied Science, Harvard University, Cambridge, Massachusetts 02138, USA}

\author{Roy J. Garcia}
\email{roygarcia@g.harvard.edu}
\affiliation{Department of Physics, Harvard University, Cambridge, Massachusetts 02138, USA}

\author{Kaifeng Bu}
\email{kfbu@fas.harvard.edu}
\affiliation{Department of Physics, Harvard University, Cambridge, Massachusetts 02138, USA}

\author{Arthur Jaffe}
\email{jaffe@g.harvard.edu}
\affiliation{Department of Physics, Harvard University, Cambridge, Massachusetts 02138, USA}

\date{\today}

\begin{abstract}
\textit{Magic}, or nonstabilizerness, characterizes how far away a state is from the stabilizer states, making it an important resource in quantum computing, under the formalism of the Gotteman-Knill theorem. 
In this paper, we study the magic of the $1$-dimensional Random Matrix Product States (RMPSs) using the $L_{1}$-norm measure. We firstly relate the $L_{1}$-norm to the $L_{4}$-norm. We then employ a unitary $4$-design to map the $L_{4}$-norm to a $24$-component statistical physics model. By evaluating partition functions of the model, we obtain a lower bound on the expectation values of the $L_{1}$-norm. This bound grows exponentially with respect to the qudit number $n$, indicating that the $1$D RMPS is highly magical. Our numerical results confirm that the magic grows exponentially in the qubit case.

\end{abstract}

\maketitle

\section{Introduction}
Quantum resources~\cite{QRT_RMP_2018}, including entanglement~\cite{Wootters_1998_Entanglement_Resource,Horodechi_PRA_Entanglement}, coherence~\cite{Aberg_2006_Coherence,Aberg_2014_Coherence_PRL}, magic~\cite{Kitaev_magic_state_distillation}, and uncomplexity~\cite{Brown_Quantum_complexity,Halperin_Uncomplexity_2021}, play a crucial role in understanding various quantum effects. 
Among these resources, magic, a quantity which characterizes the distance between a state (gate) and the stabilizer states (Clifford gates), has been proposed as a resource in quantum computation
\cite{Gottesman-Knill_theorem,nest2010classical,jozsa2014classical, koh2017further, bouland2018complexity,yoganathan2019quantum}. In recent years, various measures of magic have been introduced  to quantify the amount of magic in quantum states and circuits~\cite{Veitch12mag,Veitch14,bravyi2016trading,bravyi2019simulation,howard2017application,seddon2021quantifying,
bu2019efficient,seddon2019quantifying,wang2019quantifying,KBu2022QuantumML,KBu2021QuantumML2,Leone2022,haug2023stabilizer,haug2023efficient,Haug23,BGJ23a,BGJ23b,BGJ23c}. They have also been used to bound classical
simulation times in quantum computation
\cite{bravyi2016trading,bravyi2019simulation,howard2017application,seddon2021quantifying,
bu2019efficient,bu2022classical,seddon2019quantifying,wang2019quantifying}. The connection between magic and statistical complexity in theoretical machine learning enables it to characterize the capacity of quantum neural networks~\cite{KBu2022QuantumML,KBu2021QuantumML2}.

To realize quantum computation on a number of qubits, one has to prepare quantum states with plentiful quantum resources in a quantum many-body system. A feasible way to do this is to exploit the states emergent from the ground states of a gapped Hamiltonian with finite range interactions via a cooling procedure~\cite{Entanglement_RMPS}. The \textit{matrix product states} (MPSs), a type of tensor network, are powerful in studying the ground states of gapped one-dimensional many-body Hamiltonians, of which the $AKLT$ ground state is a paradigmatic example~\cite{AKLT_model}. The ground states and some low-lying excited states of many low-dimensional quantum many-body systems can be approximated by MPSs, which is also known as the density-matrix renormalization group (DMRG)~\cite{Schollwock2005DMRG,Farhi2008DMRG_TFIM,NAKATANI2018DMRG}. 
Recently, the magic of quantum  many-body  states has been studied \cite{Liu_2022,Magic_of_MPS,Oliviero_2022_TFIM_Magic},  including the translationally-invariant (TI) MPSs~\cite{Magic_of_MPS}. The authors propose an efficient way to calculate the Stabilizer R\'{e}nyi Entropy \cite{Leone2022},
a magic monotone, and show that the magic of the ground state of a 1D Transverse Field Ising Model is extensive. 

Random matrix product states (RMPSs), a random version of MPSs, have been used as a tool 
to study properties of many-body system, such as  statistical properties, correlations and entanglement ~\cite{RMPS_statistical_properties,RMPS_correlation,Entanglement_RMPS}.
Recently, the RMPS  has also been proved to play an important role in overcoming barren plateaus arising in quantum machine learning~\cite{Liu_2021_RMPS_QML,Garcia_2023}. Moreover, since the RMPS is typical in the phase of quantum matter, one can employ it to approximate the ground states of some general disordered parent Hamiltonians~\cite{Schuch2011_MPS_classification}.  
One can also understand whether such states provide ample quantum resources for quantum computation applications through the framework of the resource theory of magic~\cite{Veitch14}.

We  obtain a lower bound on the magic of RMPSs, in terms of an $L_4$-norm.  In more detail, we bound the $L_{1}$-norm measure of magic for a  RMPS, and transform the problem to a calculation of the $L_{4}$-norm. We employ a unitary 4-design to map it to a 24-component spin model in statistical mechanics. By calculating partition functions of the model with different nearest neighbour interactions, we obtain an upper bound on the $L_{4}$-norm, and establish a lower bound on the magic of an RMPS. 

We find that with a high probability, RMPSs have  an exponentially large magic with respect to the system size $n$. Therefore, it is possible to experimentally prepare such states from a disordered parent Hamiltonian with an ample amount of quantum resource. In parallel with previous efforts \cite{Magic_of_MPS,Oliviero_2022_TFIM_Magic}, our work provides another perspective to understand the nonstabilizerness of many-body systems based on the typicality, and we obtain the consistent extensive magic of the many-body states \cite{Magic_of_MPS}. From a quantum information perspective, one would expect that a state is resourceful when it is capable of accomplishing some quantum computational tasks. The RMPS is a model used in quantum machine learning; measuring its magic provides a step in understanding its learning capabilities~\cite{Liu_2021_RMPS_QML,Garcia_2023,KBu2022QuantumML,KBu2021QuantumML2}.

We organize our presentation as follows: in Sec.~\ref{sec:The main result}, we investigate the magic of a 1-dimensional RMPS, where we employ a technique used in previous studies~\cite{RMPS_partition_function,Entanglement_RMPS} to map the local expectation values of unitary designs to the calculation of partition functions of the statistical physics model. This produces an exponentially large lower bound on the magic. In Sec.~\ref{sec:Numerical results}, we present numerical calculations for the magic of RMPSs composed of qubits, which grows exponentially with the system size, consistent with our theoretical prediction. In Sec.~\ref{sec:The summary and outlook}, we briefly summarize our work and propose some further directions of study.

\section{Main Results}\label{sec:The main result}
In this section, we provide definitions for the RMPS and the magic monotone used in this work. We state our main bound on the magic of an RMPS in Theorem~\ref{thm: Theorem 1}. We establish this result by assuming a bound \eqref{eqn: Upper bound on 4-th moment summation} and later proving it.

\subsection{Magic}
For an $n$-qudit system with local dimension $d$ in Hilbert space $\mathscr{H}=(\mathbb{C}^{d})^{\otimes n}$, the generalized Pauli group is 
\begin{equation} \label{eqn:generalized Pauli}
    \mathcal{P}^{n} = \{ P_{\vec{a}}:P_{\vec{a}}\equiv\otimes_{i}P_{a_{i}} \}_{\vec{a}\in V^{n}}\;, 
\end{equation}
where $P_{a_{i}}= X^{r_{i}}Z^{s_{i}}$ for any $a_{i} = (r_{i},s_{i})\in V \equiv \mathbb{Z}^{d} \times \mathbb{Z}^{d}$ and $\vec{a}=(a_1,\ldots,a_n)$. The qudit Pauli $X$ and $Z$ operators are defined by $X\ket{j} = \ket{j+1\ {\rm mod}\ d},Z\ket{j} = \exp(i\frac{2\pi j}{d})\ket{j}$. 
Clifford unitaries are defined to map the Pauli group to itself. The Clifford group is ${Cl_{n}=\{U\in U(d^n):UPU^{\dagger}\in \mathcal{P}^n,\forall P \in \mathcal{P}^n\}}$. 
The set of stabilizer states is composed of states generated by the action of a Clifford unitary on $\ket{0}^{\otimes n}$,
${\text{STAB}:=\{U\ket{0}^{\otimes n}:U\in Cl_n\}}$.


The $L_{1}$-norm is a magic monotone for an $n$-qudit quantum state $\ket{\psi}$ and is defined as
\begin{equation} \label{eqn:magic measure}
    M(\ket{\psi}) \equiv \frac{1}{d^{n}}\sum_{\vec{a}\in V^{n}} \left|{\rm tr}[P_{\vec{a}}\ket{\psi}\bra{\psi}]\right| = \frac{1}{d^{n}}\sum_{\vec{a}\in V^{n}} \left|{\rm tr}[P_{\vec{a}}\rho_{\psi}]\right|\;.
\end{equation}
The $L_1$-norm is also known as the $1/2$-quantum Fourier R\'enyi entropy (see Bu et al.~\cite{Bu_2022_complexity}) and  the $1/2$-stabilizer R\'enyi entropy (see Leone et al.~\cite{Leone2022}). This monotone is \textit{faithful}, i.e. $M(\ket{\psi})=1$ if and only if $\ket{\psi}\in \text{STAB}$, and satisfies $M(\ket{\psi})>1$ for a non-stabilizer state $\ket{\psi}$. Moreover, it is \textit{stable} under free operations $U\in Cl_n$, i.e., $M(U\ket{\psi}) = M(\ket{\psi})$. This quantity was firstly proposed by Rall et al.~\cite{Rall_2019_Pauli_Propagation}. It is operationally meaningful in that it bounds the simulation cost of a quantum circuit; namely, it lower bounds the sample complexity of the Pauli propagation algorithm. It is also meaningful in that it bounds \textit{Robustness of Magic} (ROM)~\cite{haug2023stabilizer}, a well-known magic measure which bounds the classical simulation overhead of a quantum circuit via a Gottesman-Knill-type scheme \cite{Howard_2017_Robustness_of_Magic}. Especially, the ROM of magic states like $\ket{T}^{\otimes n}$ is exponentially large. The ROM also quantifies the maximal advantage attainable by resource states in some subchannel discrimination problems ~\cite{Bu2017ROM}. Since calculating ROM is an intractable optimization problem in an exponentially large space, this bound provides an efficient way to obtain ROM. The 1/2-stabilizer R\'{e}nyi entropy provides a lower bound on the number of $T$ gates (a resource gate) $t(U)$ required to implement a certain unitary $U$ by Clifford+$T$ gate sets, known as the ``$T$ count'' \cite{Jiang_2023,Leone_2023_Estimation_Fidelity}.

We prove in Appendix \ref{appendix:derivation of 4 moment inequality} that the magic $M(\ket{\psi})$ satisfies
\begin{equation} \label{eqn:inequality for magic}
    M(\ket{\psi}) \geq \frac{d^{n/2}}{(\sum_{\vec{a}}{\rm tr}[(P_{\vec{a}}\ket{\psi}\bra{\psi})^{\otimes 4}])^{1/2}}\;,
\end{equation}
where the sum is taken over $V^n$. Therefore, one can obtain a lower bound on $M(\ket{\psi})$ by evaluating the sum over the 4-th moment of Pauli operators $\sum_{\vec{a}}{\rm tr}[(P_{\vec{a}}\ket{\psi}\bra{\psi})^{\otimes 4}]$. Since the Clifford group is a unitary 3-design \cite{Webb_Clifford_3_design,Zhu_2017_Clifford_3_design}, namely the first three moments of the average over Haar random unitaries can be well approximated by random Clifford unitaries, the 4-th moment is the lowest nontrivial moment one should use to distinguish the Haar random ensemble from the Clifford ensemble.

\subsection{The Random Matrix Product States}
A Matrix Product State on $n$ qudits is defined as
\begin{equation} \label{eqn:definition of MPS}
    \ket{\psi} = \sum_{i_j}{\rm tr}\left[ A^{i_{1}}_{1}A^{i_{2}}_{2}\cdots A^{i_{n}}_{n} \right]\ket{i_{1} i_{2} \cdots i_{n}},
\end{equation}
where $A^{i_{j}}$ are $B \times B$ matrices, and $B$ denotes the \textit{bond dimension}. In \eqref{eqn:definition of MPS}, the $i_{1},\cdots,i_{n}$ are spin indices, with values $0,\cdots,d-1$, to be contracted with the basis states $\ket{i_{1},\cdots,i_{n}}$, where $d$ is the local dimension. Since each $A$ is a tensor with two indices, the MPS is represented graphically as
\begin{equation}
\begin{tikzpicture}[baseline={([yshift=-.5ex]current bounding box.center)},inner sep=1mm]
   \foreach \i in {1,...,5} {
        \node[tensor_blue] (\i) at (\i, 0) {$A_\i$};
        \node (\i spin) at (\i, -0.7) {$i_\i$};
        \draw[-] (\i) -- (\i,-0.5);
    };
    \foreach \i in {1,...,4} {
        \pgfmathtruncatemacro{\iplusone}{\i + 1};
        \draw[-] (\i) -- (\iplusone);
    };
    \draw(5) -- +(0.5,0) node[right] (dot) {$\cdots$};

    \node[tensor_blue] (n) at (6.7,0) {$A_{n}$};
    \node (n spin) at (6.7, -0.7) {$i_n$};
    \draw[-] (n) -- (6.7,-0.5);
     \draw (n) -- (dot);
     \draw[-] (1.west) .. controls +(-1.5, 1) and +(1.5, 1) .. (n.east);
\end{tikzpicture}
\end{equation}
where the bond indices are contracted. If the MPSs can be unitarily embedded \cite{PRA_Unitarily_embedded_MPS,Unitarily_embedded_MPS}, each tensor is equipped with another leg connected with a state vector $\ket{0}\in \mathbb{C}^{d}$
\begin{equation}\label{eqn:RMPS}
\ket{\psi}=
\begin{tikzpicture}[baseline={([yshift=-.5ex]current bounding box.center)},inner sep=1mm]
   \foreach \i in {1,...,5} {
        \node[tensor_blue] (\i) at (\i, 0) {$U_\i$};
        \node (\i spin) at (\i, -0.7) {$i_\i$};
        \node[tensor_green] (0\i) at (\i,0.7) {$\ket{0}$};
        \draw[-] (\i) -- (\i,-0.5);
        \draw[-] (\i) -- (0\i);
    };
    \foreach \i in {1,...,4} {
        \pgfmathtruncatemacro{\iplusone}{\i + 1};
        \draw[-] (\i) -- (\iplusone);
    };
    \draw(5) -- +(0.5,0) node[right] (dot) {$\cdots$};

    \node[tensor_blue] (n) at (6.7,0) {$U_{n}$};
    \node (n spin) at (6.7, -0.7) {$i_n$};
    \node[tensor_green] (0n) at (6.7,0.7) {$\ket{0}$};
    \draw[-] (n) -- (6.7,-0.5);
    \draw (n) -- (dot);
    \draw[-] (n) -- (0n);
    \draw[-,dashed](n) -- +(0.8,0);
    \draw[-,dashed](1) -- 
    +(-0.8,0);
\end{tikzpicture}
\end{equation}
where the dashed line represents periodic boundary conditions, and $U_1,\cdots,U_{n}\in U(dB)$ are unitaries mapping the input from $\mathbb{C}^{d}\otimes \mathbb{C}^{B}$ to the output in $\mathbb{C}^{d}\otimes \mathbb{C}^{B}$. When $U_1,\cdots,U_{n}$ are i.i.d. Haar random unitaries sampled from the unitary group, the MPS is called a \textit{Random Matrix Product State} (RMPS), whose norm is proven to have exponential contraction to $1$ in the large $n$ limit~\cite{Entanglement_RMPS}. We assume the large $n$ limit in this work so that $|\bra{\psi}\ket{\psi}|^2 = 1$. The measure is denoted by $\mu_{d,n,B}$, or more concisely $\mu$.

If we introduce the notation $\overline{\ket{\psi}} \equiv \ket{\psi}^{*}$, i.e. the complex conjugate of $\ket{\psi}$, the RMPS in \eqref{eqn:RMPS} produces
\begin{flalign} \label{eqn:tensor_4_RMPS}
&\ket{\psi}^{\otimes 4} \otimes \overline{\ket{\psi}}^{\otimes 4} = \notag\\
&\begin{tikzpicture}[baseline={([yshift=-.5ex]current bounding box.center)},inner sep=1mm]
    \foreach \j in {7,6,5,4} {  
        \foreach \i in {1,...,5} {
        \node[tensor_blue] (\i) at (\i+\j*0.15, 0+\j*0.2) {$\overline{U_\i}$};
        \node[tensor_green] (0\i) at (\i+\j*0.15,0.9+\j*0.2) {$\overline{\ket{0}}$};
        \draw[-] (\i) -- +(0,-0.5);
        \draw[-] (\i) -- (0\i);
        }
        \foreach \i in {1,...,4} {
        \pgfmathtruncatemacro{\iplusone}{\i + 1};
        \draw[-] (\i) -- (\iplusone);
        };
        \draw[-,dashed](5) -- +(0.8,0);
        \draw[-,dashed](1) -- 
        +(-0.8,0);
 };
 \foreach \j in {3,2,...,0} {  
        \foreach \i in {1,...,5} {
        \node[tensor_blue] (\i) at (\i+\j*0.15, 0+\j*0.2) {$U_\i$};
        \node[tensor_green] (0\i) at (\i+\j*0.15,0.9+\j*0.2) {$\ket{0}$};
        \draw[-] (\i) -- +(0,-0.5);
        \draw[-] (\i) -- (0\i);
        }
        \foreach \i in {1,...,4} {
        \pgfmathtruncatemacro{\iplusone}{\i + 1};
        \draw[-] (\i) -- (\iplusone);
        };
        \draw[-,dashed](5) -- +(0.8,0);
        \draw[-,dashed](1) -- 
        +(-0.8,0);
 }
\end{tikzpicture}
\end{flalign}
where the former and latter four RMPSs correspond to $\ket{\psi}^{\otimes 4}$ and $\overline{\ket{\psi}}^{\otimes 4}$, respectively. 

The average over $U^{\otimes 4}\otimes \overline{U}^{\otimes 4}$ is evaluated using the Weingarten calculus~\cite{Integration_Haar_unitary_1,Integration_Haar_Unitary_2,Weingarten_func_S4}
\begin{equation} \label{eqn:t-design Weingarten calculus}
    \mathbb{E}_{U\sim \mu} U^{\otimes 4} \otimes \overline{U}^{\otimes 4} = \sum_{\sigma,\pi \in S_{4}} {\rm Wg}(\sigma^{-1}\pi , q)\ket{\sigma}\bra{\pi},
\end{equation}
where ${\rm Wg}(\sigma^{-1}\pi,q)$ is the Weingarten function defined in terms of the local dimension $q$ (in our case, $q=dB$) and $\sigma,\pi$ are permutations in the permutation group $S_{4}$ on $(\mathbb{C}^{q})^{\otimes 4}$. We define the state $\ket{\sigma}= (\mathbb{I}\otimes r(\sigma))\ket{\Omega}$, where $r$ is the representation of $S_{4}$, and $\ket{\Omega} = \sum_{j=1}^{q^{t}}\ket{j,j}$ is the maximally entangled state vector (see Appendix \ref{appendix:The Weingarten calculus and inner products} for further discussion on the notation). The main result of this work is the following theorem.
\begin{theorem}[\bf Magic of RMPSs] \label{thm: Theorem 1} Let $\ket{\psi}$ be an RMPS drawn from $\mu_{d,n,B}$. Then the magic of $\ket{\psi}$ grows exponentially with respect to the system size $n$ with overwhelming probability:
\begin{equation} \label{eqn:Markov's inequality for magic}
    {\rm Pr}\left(\log_{d} M(\ket{\psi})\geq \Omega(n)\right) \geq 1-e^{-\Omega(n)}\;.
\end{equation}
\end{theorem}

We remark that the widely utilized $1/2$-stabilizer R\'{e}nyi entropy, discussed in Leone et al.~\cite{Leone2022} and subsequently explored in the context of Matrix Product States (MPSs)~\cite{Magic_of_MPS}, is congruent to the logarithm of our $L_1$-norm measure, up to an additive constant. Therefore, our result implies that the ``$T$-count'' of an RMPS $\ket{\psi}$ grows linearly with respect to $n$ with high probability; this saturates the upper bound for any MPS~\cite{Cirac_2021_MPS_PEPS}. Moreover, since the $L_1$-norm measure lower bounds the Robustness of Magic (i.e. $\mathcal{D}(\rho)\leq \mathcal{R}(\rho)$ in~\cite{Howard_2017_Robustness_of_Magic,Rall_2019_Pauli_Propagation}), and the latter is exponentially large for some magic states, e.g. $\ket{T}^{\otimes n}$, we can conclude that an RMPS is highly probable to contain a large amount of magic. See Section~\ref{sec:Numerical results} for more details.

Let us assume an upper bound for the expectation value of summing over the 4-th moments of Pauli operators,
\begin{equation} \label{eqn: Upper bound on 4-th moment summation}
    \sum_{\vec{a}}\mathbb{E}\ {\rm tr}[(P_{\vec{a}} \ket{\psi}\bra{\psi})^{\otimes 4}]\leq C^{n}\;,
    \quad\text{with} \quad C<d\;.
\end{equation}
By Markov's inequality 
\begin{equation}
    {\rm Pr}(X\geq aE(X))\leq \frac{1}{a}\;,
\end{equation}
where $E(X)$ is the expectation value of a random variable $X$, and $a>0$. Then using \eqref{eqn:inequality for magic}
\begin{equation}
    {\rm Pr}\left(M(\ket{\psi})\leq\left(\frac{d}{C}\right)^{n/2}\frac{1}{\sqrt{a}}\right)\leq \frac{1}{a}\;.
\end{equation}
Once $C<d$, we can pick a positive number $0<c_{1}<\log(d/C)$ and let $a=e^{c_{1}n}$, to obtain
\begin{equation}
    {\rm Pr}\left(M(\ket{\psi})\leq e^{\frac{1}{2}(\log(d/C)-c_{1})n}\right) \leq e^{-c_{1}n}\;,
\end{equation}
which gives us
\begin{equation}
    {\rm Pr}\left(M(\ket{\psi})\geq e^{\Omega(n)}\right) \geq 1-e^{-\Omega(n)}\;.
\end{equation}
By taking the log, we obtain \eqref{eqn:Markov's inequality for magic}. 

The remainder of this paper (including the appendices) is a proof of the upper bound in \eqref{eqn: Upper bound on 4-th moment summation}, which completes the proof of Theorem~\ref{thm: Theorem 1}. 

\subsection{The partition functions and interaction blocks}\label{sec:partition functions and interactions}
If we introduce the notation $\ket{\psi}^{\otimes 4,4} \equiv \ket{\psi}^{\otimes 4}\otimes \overline{\ket{\psi}}^{\otimes 4}$, by employing the Weingarten calculus introduced in ~\eqref{eqn:t-design Weingarten calculus} and Table \ref{tab:S4 Weingarten function}, the expectation value of the 4-th moment ${\rm tr}\left[(P_{\vec{a}} \ket{\psi}\bra{\psi})^{\otimes 4}\right]$ is
\begin{flalign} \label{eqn:expectation of Pa}
    &\mathbb{E}_{\psi \sim \mu}{\rm tr}\left[(P_{\vec{a}} \ket{\psi}\bra{\psi})^{\otimes 4}  \right]=  \notag\\
    &\sum_{\{S_{4}\}^{2n}}
\begin{tikzpicture}[baseline={([yshift=-.5ex]current bounding box.center)},inner sep=1mm]
\tikzset{decoration={snake,amplitude=.5mm,segment length=2mm,post length=0mm,pre length=0mm},
Paulistring/.style = {rectangle, draw=black, thick, fill=blue!25, minimum width=7.2cm,minimum height = 0.6cm},}
\node[Paulistring] (P) at (4.5,-1.7) {$P_{\vec{a}}$};
 \foreach \i in {1,...,5} {
        \node [fill=black,circle,inner sep=0pt, minimum size=2mm] (u\i) at (1.5*\i,0) {};
        \node [fill=black,circle,inner sep=0pt, minimum size=2mm] (d\i) at (1.5*\i,-1) {};
        \node[tensor_green] (0\i) at (1.5*\i,0.7) {$\ket{0}^{\otimes 4,4}$};
        \draw[decorate,thick] (u\i) -- (d\i);
        \draw[-,draw=red] (u\i) -- (0\i);
        \draw[-,draw=red] (d\i) -- (d\i.south|-P.north);
        };   
        \foreach \i in {1,...,4} {
        \pgfmathtruncatemacro{\iplusone}{\i + 1};
        \draw[-,draw=blue] (d\i) -- (u\iplusone);
        };
        \draw[-,dashed](d5) -- +(0.6,0.4);
        \draw[-,dashed](u1) -- 
        +(-0.6,-0.4);
\end{tikzpicture}
\end{flalign}
where the $P_{\vec{a}}$ block is the shorthand notation for the operator $(\mathbb{I}\otimes P_{\vec{a}})^{\otimes 4}$, and the black dots represent one of the elements in $S_{4}$, so the summation $\{S_{4}\}^{2n}$ is over all configurations of the $2n$ black dots. The wavy lines are the Weingarten functions ${\rm Wg}(\sigma^{-1}\pi,q)$ for each pair of black dots with permutation ($\sigma,\pi$). The red lines and blue lines are contractions over $\mathbb{C}^{d}$ and $\mathbb{C}^{B}$, respectively. In \eqref{eqn:expectation of Pa}, the $n$ contractions between permutations and the state $\ket{0}^{\otimes 4,4}$ are
\begin{equation}
    \bra{\pi}\ket{0}^{\otimes4,4} = \bra{0}\ket{0}^{4} = 1\;.
\end{equation}
 
Since $P_{\vec{a}}\in\{I,X,Z,XZ,\cdots,X^{d-1},Z^{d-1}\}^n$, we can decompose the $P_{\vec{a}}$ into local operators at each site, so for a randomly chosen $P_{\vec{a}}$, \eqref{eqn:expectation of Pa} becomes (where we omit the subscript of $\mathbb{E}$)
\begin{flalign} \label{eqn:expectation of Pa local operators}
    &\mathbb{E}\ {\rm tr}[(P_{\vec{a}} \ket{\psi}\bra{\psi})^{\otimes 4}] =  \notag\\
    &\sum_{\{S_{4}\}^{2n}}
\begin{tikzpicture}[baseline={([yshift=-.5ex]current bounding box.center)},inner sep=1mm]
\tikzset{decoration={snake,amplitude=.5mm,segment length=2mm,post length=0mm,pre length=0mm},
Paulistring/.style = {rectangle, draw=black, thick, fill=blue!25, minimum width=7.2cm,minimum height = 0.6cm},}
 \foreach \i in {1,...,5} {
        \node [fill=black,circle,inner sep=0pt, minimum size=2mm] (u\i) at (1.2*\i,0) {};
        \node [fill=black,circle,inner sep=0pt, minimum size=2mm] (d\i) at (1.2*\i,-1) {};
        \ifnum \i=1
        \node[tensor_green] (O\i) at (1.2*\i,-1.7) {$\ket{I}$};
        \else \ifnum \i>3
                \node[tensor_purple] (O\i) at (1.2*\i,-1.7) {$\ket{O_2}$};
                \else
                \node[tensor_blue] (O\i) at (1.2*\i,-1.7) {$\ket{O_1}$};
                \fi
        \fi
        \draw[decorate,thick] (u\i) -- (d\i);
        \draw[-,draw=red] (d\i) -- (O\i.north);
        }
        \foreach \i in {1,...,4} {
        \pgfmathtruncatemacro{\iplusone}{\i + 1};
        \draw[-,draw=blue] (d\i) -- (u\iplusone);
        };
        \draw[-,dashed](d5) -- +(0.6,0.4);
        \draw[-,dashed](u1) -- 
        +(-0.6,-0.4);
\end{tikzpicture}
\end{flalign}
where $\ket{I} = (\mathbb{I}\otimes \mathbb{I})\ket{\Omega},\ket{O_1} = (\mathbb{I}\otimes O_1)\ket{\Omega},\ket{O_2}=(\mathbb{I}\otimes O_2)\ket{\Omega}$, with $\mathbb{I},O_1,O_2$ acting on $(\mathbb{C}^{d})^{\otimes 4}$. Here we distinguish the two nonidentity operators $O_1$ and $O_2$ by their different contractions with permutations $\sigma\in S_{4}$. For an operator $O$, the contractions $\bra{\sigma}\ket{O}$ are summarized in Table \ref{tab:S4 operator inner product} and discussed in Appendix \ref{appendix:The Weingarten calculus and inner products}. If we write the contraction $\bra{\sigma}\ket{O}$ of a Pauli operator $O$ in terms of the five rows in Table \ref{tab:S4 operator inner product} into a vector, we have
\begin{equation}
    \bra{\sigma}\ket{O}=(0,0,a,0,b),
\end{equation}
with $a=d^2$ or $0$ and $b=d$ or $0$ for different local dimensions. There are three different cases: \RN{1}. $d$ is odd, \RN{2}. $d=2k$ for odd $k$, \RN{3}. $d=4k$ for integer $k$.

\RN{1}. \textit{$d$ is odd}. When $d$ is odd, for any $l\neq 0\  ({\rm mod}\ d),m\neq 0\  ({\rm mod}\ d)$, the square and fourth power of a Pauli operator $O = X^{l}Z^{m}$, namely $O^2=X^{2l}Z^{2m}$ and $O^4=X^{4l}Z^{4m}$ (with unimportant phase factors) are always nonidentity operators, so ${\rm tr}O^2=0$ and ${\rm tr}O^4=0$. We have
\begin{equation} \label{eqn:contraction in odd d}
    \bra{\sigma}\ket{O} = (0,0,0,0,0)\;,
\end{equation}
for all nonidentity operators.

\RN{2}. \textit{$d=2k$, for odd $k$}. In this case, only the Pauli operators $O=X^{k},Z^{k},X^{k}Z^{k}$ have $O^2=I$ and $O^4=I$, so the contraction $\bra{\sigma}\ket{O}$ is
\begin{equation} \label{eqn:contraction in d=2k}
    \bra{\sigma}\ket{O} = 
\left\{  
    \begin{array}{ll}  
        (0,0,d^2,0,d), & O=X^{k},Z^{k}\ {\rm or}\ X^{k}Z^{k}\;, \\
        (0,0,0,0,0)\;, & {\rm otherwise.}
        \end{array}  
\right.
\end{equation}

\RN{3}. \textit{$d=4k$, for integer $k$}. In this case, the Pauli operators $O=X^{k},Z^{k},X^{k}Z^{k}$ satisfy $O^2\neq I$ and $O^4=I$, and the operators $O=X^{2k},Z^{2k},X^{2k}Z^{2k}$ satisfy $O^2=I$ and $O^4=I$. All other operators have a non-identity square and fourth power, so the contraction $\bra{\sigma}\ket{O}$ is
\begin{equation} \label{eqn:contraction in d=4k}
    \bra{\sigma}\ket{O} = 
\left\{  
    \begin{array}{ll}  
        (0,0,0,0,d)\;, & O=X^{k},Z^{k}\ {\rm or}\ X^{k}Z^{k}\;, \\
        (0,0,d^2,0,d)\;, & O=X^{2k},Z^{2k}\ {\rm or}\ X^{2k}Z^{2k}\;,\\
        (0,0,0,0,0)\;, & {\rm otherwise.}
        \end{array}  
\right.
\end{equation}

Therefore, from the three cases, we can classify the Pauli operators with nonzero contraction into two types: $O_{1}$ with $\bra{\sigma}\ket{O_{1}}=(0,0,d^2,0,d)$, and $O_{2}$ with $\bra{\sigma}\ket{O_{2}}=(0,0,0,0,d)$. The blue lines in ~\eqref{eqn:expectation of Pa local operators} are contractions between two different permutations $\sigma,\pi\in S_{4}$, which are listed in Table \ref{tab:S4 inner product}, where the calculation is simply checking the number of closed permutations in $s=\sigma^{-1}\pi$, which is the corresponding power of $q$.

Combining all of the results, we can define the following interaction blocks~\cite{Entanglement_RMPS}
\begin{equation}\label{eqn:the Identity Block}
    \begin{tikzpicture}[baseline={([yshift=-.5ex]current bounding box.center)},inner sep=1mm]
        \node[tensor_green] (I) at (0,0) {};
        \node[inner sep=0pt, minimum size=2mm] (left) at (-0.8,0) {$\sigma$};
        \node[inner sep=0pt, minimum size=2mm] (right) at (0.8,0) {$\pi$};
        \draw[-](I) -- (left);
        \draw[-](I) -- (right);
    \end{tikzpicture}
    =\sum_{\{ S_{4}\}}
\begin{tikzpicture}[baseline={([yshift=-.5ex]current bounding box.center)},inner sep=1mm]
    \tikzset{decoration={snake,amplitude=.5mm,segment length=2mm,post length=0mm,pre length=0mm},
Paulistring/.style = {rectangle, draw=black, thick, fill=blue!25, minimum width=7.2cm,minimum height = 0.6cm},}
    \node [inner sep=0pt, minimum size=2mm] (u1) at (1.2,0) {$\sigma$};
    \node [inner sep=0pt, minimum size=2mm] (u2) at (1.2*2,0) {$\pi$};
    \node [fill=black,circle,inner sep=0pt, minimum size=2mm] (d1) at (1.2,-1) {};
    \node[tensor_green] (I1) at (1.2,-1.7) {$\ket{I}$};
    \draw[decorate,thick] (u1) -- (d1);
    \draw[-,draw=red] (d1) -- (I1.north);
    \draw[-,draw=blue] (d1) -- (u2);
    \end{tikzpicture}
\end{equation}

\begin{equation}\label{eqn:the O1 Block}
    \begin{tikzpicture}[baseline={([yshift=-.5ex]current bounding box.center)},inner sep=1mm]
        \node[tensor_blue] (O) at (0,0) {};
        \node[inner sep=0pt, minimum size=2mm] (left) at (-0.8,0) {$\sigma$};
        \node[inner sep=0pt, minimum size=2mm] (right) at (0.8,0) {$\pi$};
        \draw[-](O) -- (left);
        \draw[-](O) -- (right);
    \end{tikzpicture}
    =\sum_{\{ S_{4}\}}
\begin{tikzpicture}[baseline={([yshift=-.5ex]current bounding box.center)},inner sep=1mm]
    \tikzset{decoration={snake,amplitude=.5mm,segment length=2mm,post length=0mm,pre length=0mm},
Paulistring/.style = {rectangle, draw=black, thick, fill=blue!25, minimum width=7.2cm,minimum height = 0.6cm},}
    \node [inner sep=0pt, minimum size=2mm] (u1) at (1.2,0) {$\sigma$};
    \node [inner sep=0pt, minimum size=2mm] (u2) at (1.2*2,0) {$\pi$};
    \node [fill=black,circle,inner sep=0pt, minimum size=2mm] (d1) at (1.2,-1) {};
    \node[tensor_blue] (O1) at (1.2,-1.7) {$\ket{O_1}$};
    \draw[decorate,thick] (u1) -- (d1);
    \draw[-,draw=red] (d1) -- (O1.north);
    \draw[-,draw=blue] (d1) -- (u2);
    \end{tikzpicture}
\end{equation}

\begin{equation}\label{eqn:the O2 Block}
    \begin{tikzpicture}[baseline={([yshift=-.5ex]current bounding box.center)},inner sep=1mm]
        \node[tensor_purple] (O) at (0,0) {};
        \node[inner sep=0pt, minimum size=2mm] (left) at (-0.8,0) {$\sigma$};
        \node[inner sep=0pt, minimum size=2mm] (right) at (0.8,0) {$\pi$};
        \draw[-](O) -- (left);
        \draw[-](O) -- (right);
    \end{tikzpicture}
    =\sum_{\{ S_{4}\}}
\begin{tikzpicture}[baseline={([yshift=-.5ex]current bounding box.center)},inner sep=1mm]
    \tikzset{decoration={snake,amplitude=.5mm,segment length=2mm,post length=0mm,pre length=0mm},
Paulistring/.style = {rectangle, draw=black, thick, fill=blue!25, minimum width=7.2cm,minimum height = 0.6cm},}
    \node [inner sep=0pt, minimum size=2mm] (u1) at (1.2,0) {$\sigma$};
    \node [inner sep=0pt, minimum size=2mm] (u2) at (1.2*2,0) {$\pi$};
    \node [fill=black,circle,inner sep=0pt, minimum size=2mm] (d1) at (1.2,-1) {};
    \node[tensor_purple] (O1) at (1.2,-1.7) {$\ket{O_2}$};
    \draw[decorate,thick] (u1) -- (d1);
    \draw[-,draw=red] (d1) -- (O1.north);
    \draw[-,draw=blue] (d1) -- (u2);
    \end{tikzpicture}
\end{equation}
where the permutation $\tau$ at the black dot is summed over. The three blocks are $24\times 24$ matrices with explicit expressions given in the supplementary Mathematica notebook.

With the interaction blocks, \eqref{eqn:expectation of Pa local operators} is simplified to
\begin{flalign} \label{eqn:expectation of Pa partition function}
    &\mathbb{E}\ {\rm tr}[(P_{\vec{a}} \ket{\psi}\bra{\psi})^{\otimes 4}] =  \notag \\
    &\sum_{\{S_{4}\}^{n}}
\begin{tikzpicture}[baseline={([yshift=-.5ex]current bounding box.center)},inner sep=1mm]
\node [fill=black,circle,inner sep=0pt, minimum size=2mm] (0) at (0,0) {};
 \foreach \i in {1,...,5} {
        \node [fill=black,circle,inner sep=0pt, minimum size=2mm] (\i) at (1.2*\i,0) {};
        \ifnum \i=1
        \node[tensor_green] (O\i) at (1.2*\i-0.6,0) {};
        \else  \ifnum \i>3
                \node[tensor_purple] (O\i) at (1.2*\i-0.6,0) {};
                \else
                \node[tensor_blue] (O\i) at (1.2*\i-0.6,0) {};
                \fi
        \fi
        \draw[-] (O\i.east) -- (\i.west);
    };
    \foreach \i in {0,1,...,4} {
        \pgfmathtruncatemacro{\iplusone}{\i + 1};
        \draw[-] (\i.east) -- (O\iplusone.west);
        };
    \draw[-,dashed](5) -- +(0.6,0);
    \draw[-,dashed](0) -- 
        +(-0.6,0);
\end{tikzpicture}
\end{flalign}
which can be understood as: at each black dot (or namely at each site), there is a $24$-component spin, corresponding to the 24 group elements of $S_{4}$, and the nearest neighbour spins are interacting through the blocks. The summation over all spin configurations maps the expectation value to a partition function, as discussed in Ref.~\cite{RMPS_partition_function}. From the partition function perspective, the $24\times 24$ interaction block matrices are treated as the transfer matrices in statistical mechanics, so the expectation value in \eqref{eqn:expectation of Pa partition function} is easily obtained by taking the trace after doing matrix multiplication. The mapping simplifies the calculation of the summation $\sum_{\vec{a}}\mathbb{E}\ {\rm tr}[(P_{\vec{a}} \ket{\psi}\bra{\psi})^{\otimes 4}]$ to the summation of $d^n$ partition functions. 

\subsection{The lower bound of magic}
By explicitly solving the spectrum, we can establish an upper bound on the spectral radius $\rho$ of the interaction blocks (see Appendix \ref{appendix:more analysis on the upper bound} for details). When $d\geq 2$ and $B\geq 2$, we have
\begin{flalign}\label{eqn: The spectral radii}
\rho \begin{tikzpicture}[baseline={([yshift=-.5ex]current bounding box.center)},inner sep=1mm]
        \node[tensor_green] (I) at (0,0) {};
        \node[inner sep=0pt, minimum size=2mm,label={[xshift=0cm, yshift=-0.4cm]$($}] (left) at (-0.8,0) {};
        \node[inner sep=0pt, minimum size=2mm,label={[xshift=0cm, yshift=-0.4cm]$)$}] (right) at (0.8,0) {};
        \draw[-](I) -- (left);
        \draw[-](I) -- (right);
    \end{tikzpicture}  &\leq 1\;,\notag\\
\rho \begin{tikzpicture}[baseline={([yshift=-.5ex]current bounding box.center)},inner sep=1mm]
        \node[tensor_blue] (I) at (0,0) {};
        \node[inner sep=0pt, minimum size=2mm,label={[xshift=0cm, yshift=-0.4cm]$($}] (left) at (-0.8,0) {};
        \node[inner sep=0pt, minimum size=2mm,label={[xshift=0cm, yshift=-0.4cm]$)$}] (right) at (0.8,0) {};
        \draw[-](I) -- (left);
        \draw[-](I) -- (right);
    \end{tikzpicture} &\leq 2/d^2\;, \notag\\
    \rho \begin{tikzpicture}[baseline={([yshift=-.5ex]current bounding box.center)},inner sep=1mm]
        \node[tensor_purple] (I) at (0,0) {};
        \node[inner sep=0pt, minimum size=2mm,label={[xshift=0cm, yshift=-0.4cm]$($}] (left) at (-0.8,0) {};
        \node[inner sep=0pt, minimum size=2mm,label={[xshift=0cm, yshift=-0.4cm]$)$}] (right) at (0.8,0) {};
        \draw[-](I) -- (left);
        \draw[-](I) -- (right);
    \end{tikzpicture} &\leq 3/d^3\;.
\end{flalign}
When $d=2$ and $B\geq 2$, we have a better bound as
\begin{flalign}\label{eqn: The spectral radii at d=2}
\rho \begin{tikzpicture}[baseline={([yshift=-.5ex]current bounding box.center)},inner sep=1mm]
        \node[tensor_green] (I) at (0,0) {};
        \node[inner sep=0pt, minimum size=2mm,label={[xshift=0cm, yshift=-0.4cm]$($}] (left) at (-0.8,0) {};
        \node[inner sep=0pt, minimum size=2mm,label={[xshift=0cm, yshift=-0.4cm]$)$}] (right) at (0.8,0) {};
        \draw[-](I) -- (left);
        \draw[-](I) -- (right);
    \end{tikzpicture}  &\leq 1\;,\notag\\
\rho \begin{tikzpicture}[baseline={([yshift=-.5ex]current bounding box.center)},inner sep=1mm]
        \node[tensor_blue] (I) at (0,0) {};
        \node[inner sep=0pt, minimum size=2mm,label={[xshift=0cm, yshift=-0.4cm]$($}] (left) at (-0.8,0) {};
        \node[inner sep=0pt, minimum size=2mm,label={[xshift=0cm, yshift=-0.4cm]$)$}] (right) at (0.8,0) {};
        \draw[-](I) -- (left);
        \draw[-](I) -- (right);
    \end{tikzpicture} &\leq 1/d^2\;.
\end{flalign}
Another useful inequality is
\begin{equation} \label{eqn:Spectral radius inenquality}
    \rho(M_1+M_2)\leq \rho(M_1) + \rho(M_2),
\end{equation}
where $M_1$ and $M_2$ are square matrices with the same size. In this section, we calculate the $C$ for \eqref{eqn: Upper bound on 4-th moment summation} in the three cases in Sec.~\ref{sec:partition functions and interactions}, and we show that $C<d$ to prove Theorem~\ref{thm: Theorem 1}.
\subsubsection{The case of odd $d$}
In this case, the contraction $\bra{\sigma}\ket{O}$ is always zero for nonidentity Pauli operators $O$, as in \eqref{eqn:contraction in odd d}. Therefore, there is only one term in the summation $\sum_{\vec{a}}\mathbb{E}\ {\rm tr}[(P_{\vec{a}} \ket{\psi}\bra{\psi})^{\otimes 4}]$ when $P_{\vec{a}}=I$ as
\begin{equation}
    \mathbb{E}\ {\rm tr}[(I \ket{\psi}\bra{\psi})^{\otimes 4}] 
= {\rm tr}[ (
\begin{tikzpicture}[baseline={([yshift=-.5ex]current bounding box.center)},inner sep=0mm]
        \node[tensor_green] (I) at (0,0) {};
        \node[inner sep=0pt, minimum size=2mm] (left) at (-0.8,0) {};
        \node[inner sep=0pt, minimum size=2mm] (right) at (0.8,0) {};
        \draw[-](I) -- (left);
        \draw[-](I) -- (right);
\end{tikzpicture}
)^n ]\leq 24\;.
\end{equation}

as $\norm{A}_{1}\leq D\norm{A}_{\infty}$, where $D$ is the dimension. Therefore, for a 1D RMPS with odd local dimension $d$, we have $C=24^{1/n}<d$ in \eqref{eqn: Upper bound on 4-th moment summation}, so we have proved Theorem~\ref{thm: Theorem 1}

\subsubsection{The case $d=2k$,  for odd $k$}
In this case, the contraction $\bra{\sigma}\ket{O}$ is given in \eqref{eqn:contraction in d=2k}, so there are two types of interaction blocks: the green block for a local identity operator and the blue block for local Pauli operators $X^{k},Z^{k},X^{k}Z^{k}$. So the summation $\sum_{\vec{a}}\mathbb{E}\ {\rm tr}[(P_{\vec{a}} \ket{\psi}\bra{\psi})^{\otimes 4}]$ is obtained by taking all possibilities to put $l$ blue blocks (each contains three possibilities $X^{k},Z^{k},X^{k}Z^{k}$) over $n$ blocks as
\begin{flalign}
    &\sum_{\vec{a}} \mathbb{E}{\rm tr}[(P_{\vec{a}}\ket{\psi} \bra{\psi})^{\otimes 4}] \notag\\
&= {\rm tr}[(\begin{tikzpicture}[baseline={([yshift=-.5ex]current bounding box.center)},inner sep=0mm]
        \node[tensor_green] (O1) at (-4.8,0) {};
        \node[inner sep=0pt, minimum size=2mm] (O1left) at (-5.1,0) {};
        \node[inner sep=0pt, minimum size=2mm] (O1right) at (-4.5,0) {};
\end{tikzpicture} + 3\begin{tikzpicture}[baseline={([yshift=-.5ex]current bounding box.center)},inner sep=0mm]
        \node[tensor_blue] (O1) at (-4.8,0) {};
        \node[inner sep=0pt, minimum size=2mm] (O1left) at (-5.1,0) {};
        \node[inner sep=0pt, minimum size=2mm] (O1right) at (-4.5,0) {};
\end{tikzpicture})^n ] \leq 24\rho(\begin{tikzpicture}[baseline={([yshift=-.5ex]current bounding box.center)},inner sep=0mm]
        \node[tensor_green] (O1) at (-4.8,0) {};
        \node[inner sep=0pt, minimum size=2mm] (O1left) at (-5.1,0) {};
        \node[inner sep=0pt, minimum size=2mm] (O1right) at (-4.5,0) {};
\end{tikzpicture}+3\begin{tikzpicture}[baseline={([yshift=-.5ex]current bounding box.center)},inner sep=0mm]
        \node[tensor_blue] (O1) at (-4.8,0) {};
        \node[inner sep=0pt, minimum size=2mm] (O1left) at (-5.1,0) {};
        \node[inner sep=0pt, minimum size=2mm] (O1right) at (-4.5,0) {};
\end{tikzpicture})^{n}\;.
\end{flalign}
By \eqref{eqn:Spectral radius inenquality}, we have
\begin{flalign}
    \rho(\begin{tikzpicture}[baseline={([yshift=-.5ex]current bounding box.center)},inner sep=0mm]
        \node[tensor_green] (O1) at (-4.8,0) {};
        \node[inner sep=0pt, minimum size=2mm] (O1left) at (-5.1,0) {};
        \node[inner sep=0pt, minimum size=2mm] (O1right) at (-4.5,0) {};
\end{tikzpicture}+3\begin{tikzpicture}[baseline={([yshift=-.5ex]current bounding box.center)},inner sep=0mm]
        \node[tensor_blue] (O1) at (-4.8,0) {};
        \node[inner sep=0pt, minimum size=2mm] (O1left) at (-5.1,0) {};
        \node[inner sep=0pt, minimum size=2mm] (O1right) at (-4.5,0) {};
\end{tikzpicture}) 
&\leq \left\{  
    \begin{array}{ll}  
        1+3/d^2\;, & d=2\;, \\
        1+6/d^2\;, & d\geq 6\;.
        \end{array}  
\right.
\end{flalign}
Therefore, in this case, we have
\begin{equation}
    C=\left\{  
    \begin{array}{ll}  
        24^{1/n}(1+3/d^2)< d\;, & d=2\;, \\
        24^{1/n}(1+6/d^2)< d\;, & d\geq 6\;,
        \end{array}  
\right.
\end{equation}
which completes the proof of Theorem~\ref{thm: Theorem 1}.
When the system is composed of qubits, namely $d=2$, by \eqref{eqn:inequality for magic}, one can check that
\begin{equation} \label{eqn:qubit log magic lower bound}
   \log_{2}\mathbb{E} M(\ket{\psi}) \geq 0.1n\;,
\end{equation}
which is the lowest bound in this case.

\subsubsection{The case $d=4k$, for integer $k$}
In this case, the contraction $\bra{\sigma}\ket{O}$ in \eqref{eqn:contraction in d=4k} introduces the third purple block for local Pauli operators $X^{2k},Z^{2k},X^{2k}Z^{2k}$. The upper bound can be obtained correspondingly as
\begin{flalign}
&\sum_{\vec{a}} \mathbb{E}{\rm tr}[(P_{\vec{a}}\ket{\psi} \bra{\psi})^{\otimes 4}]\notag\\
&={\rm tr}[(\begin{tikzpicture}[baseline={([yshift=-.5ex]current bounding box.center)},inner sep=0mm]
        \node[tensor_green] (O1) at (-4.8,0) {};
        \node[inner sep=0pt, minimum size=2mm] (O1left) at (-5.1,0) {};
        \node[inner sep=0pt, minimum size=2mm] (O1right) at (-4.5,0) {};
        \end{tikzpicture}
         + 3\begin{tikzpicture}[baseline={([yshift=-.5ex]current bounding box.center)},inner sep=0mm]
        \node[tensor_blue] (O1) at (-4.8,0) {};
        \node[inner sep=0pt, minimum size=2mm] (O1left) at (-5.1,0) {};
        \node[inner sep=0pt, minimum size=2mm] (O1right) at (-4.5,0) {};
        \end{tikzpicture}
        +3\begin{tikzpicture}[baseline={([yshift=-.5ex]current bounding box.center)},inner sep=0mm]
        \node[tensor_purple] (O1) at (-4.8,0) {};
        \node[inner sep=0pt, minimum size=2mm] (O1left) at (-5.1,0) {};
        \node[inner sep=0pt, minimum size=2mm] (O1right) at (-4.5,0) {};
        \end{tikzpicture}
        )^n] 
\leq 24\left(1+\frac{9}{d^2}\right)^n\;,
\end{flalign}
where in the last line we have used $d\geq 4$ and \eqref{eqn: The spectral radii}-\eqref{eqn:Spectral radius inenquality}. Therefore, $C = 24^{1/n}(1+9/d^2)<d$ in this case, which completes the proof of Theorem~\ref{thm: Theorem 1}.
When $d=4$, one can check that the bound is
\begin{equation}
    \log_{4} \mathbb{E} M(\ket{\psi})\geq 0.34 n\;,
\end{equation}
which is the lowest bound of the expectation value of magic in this case.

\section{Numerical results}\label{sec:Numerical results}
\begin{figure}
\centering
\includegraphics[width=0.45\textwidth]{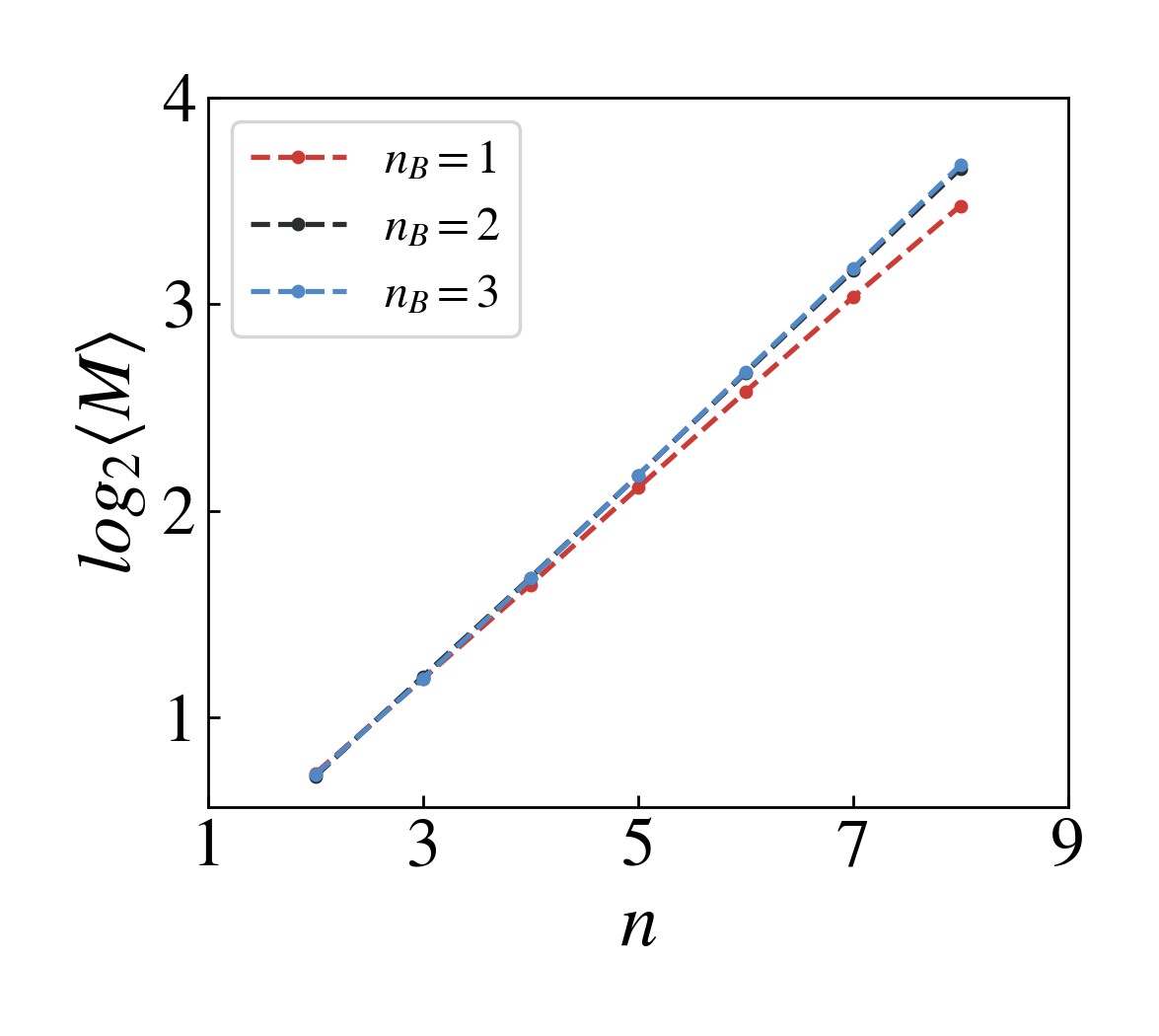}
\caption{\label{fig:The log magic in terms of n small B}The ${\rm log}_{2} (\mathbb{E}\ M(\ket{\psi}))$ in terms of $n$ for an RMPS $\ket{\psi}$, where $n=2\sim 8$, $d=2$, and $B=2,4,8$. The expectation value is sampled over $100$ RMPSs by sampling local unitaries via the Haar random measure. The linear behavior of ${\rm log}_{2} (\mathbb{E}\ M(\ket{\psi}))$ confirms the exponential growth of $\mathbb{E}\ M(\ket{\psi})$.}
\end{figure}
We numerically compute the average $\mathbb{E}\ M(\ket{\psi})$ of a $d=2$ (qubit) RMPS with bond dimensions of $B=2,4,8$. The expectation is sampled over $100$ RMPSs; namely, we generate $100$ RMPSs whose unitaries are drawn from the Haar measure on the unitary group. The logarithm of $\mathbb{E}\ M(\ket{\psi})$ versus $n$ is plotted in Figure \ref{fig:The log magic in terms of n small B}, where the linear behavior indicates that the average magic $\mathbb{E}\ M(\ket{\psi})$ grows exponentially with $n$, which is consistent with our theory. In the plot, the slope of the $B=2$ line is $\sim 0.46$, which is greater than $\sim 0.10$ in \eqref{eqn:qubit log magic lower bound}. Since $\mathcal{D}(\rho) \leq \mathcal{R}(\rho)$, and previous studies \cite{Howard_2017_Robustness_of_Magic, Heinrich_2019_ROM_Upper_Bound} have numerically shown that $\mathcal{R}(\ket{T}^{\otimes n}) \approx 1.387^n$, lower bounded by $1.366^n$, the $B=2$ line with $\mathcal{D}(\rho) \approx 1.376^n$ implies that the low bond dimension qubit RMPS is as magical as the $\ket{T}^{\otimes n}$ state. For larger $B$, the saturated $\mathcal{D}(\rho) \approx 2^{n/2} = 1.414^n$ seems to imply that the RMPS is more magical than the $\ket{T}^{\otimes n}$ state, but this may suffer from finite size effect as in \cite{Heinrich_2019_ROM_Upper_Bound}, so we leave the comprehensive analysis to future studies. This implies that, although MPSs may be efficiently simulated classically, RMPSs may still nevertheless be useful as ancilla states to introduce magic in computational problems. The details of the numerical methods are summarized in Appendix \ref{appendix:more on numerical calculation}.

\section{Summary and outlook}\label{sec:The summary and outlook}
In this paper, we explicitly calculate the magic of $1$-dimensional Random Matrix Product States. We obtain an  exponentially large lower bound 
for different local dimensions,
 suggesting the potential use of RMPSs in 
quantum computation applications such as quantum machine learning~\cite{KBu2022QuantumML,KBu2021QuantumML2,Liu_2021_RMPS_QML,Garcia_2023}. The weak dependence of the magic on the bond dimension in our theoretical study and numerical results is not a result one should expect \textit{a priori}, because entanglement and magic can generally be independent resources. This dependence differentiates RMPSs from global random Haar states and random tensor product states.

To the best of our knowledge, this is the first work to calculate the magic of a RMPS, thus exploring the typical behavior of this quantum resource of matrix product states. 
It is possible to prepare such states experimentally 
by a cooling procedure, suggesting this as a feasible way to obtain considerable quantum resources. Moreover, we believe that this study can help us understand the quantum information properties of many-body systems more clearly. People have focused on using entanglement to classify quantum phases, where two states are in the same phase if they are connected by finite-depth local unitaries (FDLU)~\cite{Wen_2010_LUT}. In a parallel manner, one can also consider using magic as a criterion for state classification, where two states are in the same phase if they are connected by Clifford unitaries. Indeed, two of the authors of this work, K. Bu and A. Jaffe, have studied that by a newly developed ``convolution group'' (CG) method to label quantum states by ``magic class''~\cite{bu2024magicclass}. Our work shows that an RMPS is a state in the highest magic class which can converge to the same CG fixed point as random product states and Haar random states.

The methods employed in this work can be applied to study random tensor networks, higher moments of Haar random unitaries and Pauli operators. For example, the entanglement entropy in measurement-induced entanglement phase transitions \cite{Bao_2020_MIPT,Iaconis_2020_MIPT,Liu_2022_MIPT} is typically obtained from evaluating the R\'{e}nyi entropies with indices $n$ and extrapolating them to the $n\to 1$ limit. By generalizing our technique, people can calculate and bound the $n$-th information theoretic quantities like conditional entropy, KL divergence by evaluating the $n$-design Weingarten calculus and bounding the resulting partition function, without going to the $B \to \infty$ limit as most of the current works do. This can help people understand the transition points at finite bond dimensions better. Furthermore, in the study of barren plateaus in the Quantum Neural Network (QNN) \cite{Liu_2021_RMPS_QML,Garcia_2023} and the Quantum Convolution Neural Network (QCNN) \cite{Pesah_2021_QCNN}, and also in the study of classical shadows~\cite{Huang_2020_Classical_Shadow,bertoni2023shallow,hu2022classical,bu2022classical_shadow}, one evaluates and bounds the expectation values of the second and third moments of Pauli operators in terms of Haar random unitaries, similar to the techniques presented here. The techniques may also be useful in computing averages of higher point out-of-time-ordered correlators (OTOCs)~\cite{Garcia_2021_higher_point_OTOC} at finite bond dimension, which can be used to study the magic of chaotic systems. We also expect that one can use these techniques to bound the sample complexity of direct fidelity estimation \cite{Leone_2023_Estimation_Fidelity}, since it is shown to be connected to the nonstabilizerness and higher point OTOCs. Recently, people have studied the entanglement entropy in higher dimensional random tensor networks such as the random Projected Entangled-Pair State (PEPS) \cite{liu2023theory}, and it is natural to apply our technique to bound the non-stabilizerness in those random tensor networks. Finally, other magic measures, such as the OTOC magic \cite{Garcia_2022_Scram}, may also be taken into consideration and studied using our method. 

We note that a recent preprint presents a counterexample showing that the 1/2-stabilizer R\'{e}nyi entropy is not a magic monotone under stabilizer measurements and Clifford operations conditioned on the measurement results \cite{haug2023stabilizer}. In our study, we focus on unitary processes without mid-circuit measurement, as in the QNN setting~\cite{Liu_2021_RMPS_QML,Garcia_2023}, in which case the $L_1$-norm measure is still a magic monotone.

\begin{acknowledgments}
We thank Ashvin Vishwanath, Boaz Barak and Leslie Valiant for comments and interesting discussion. We thank Weiyu Li for the help in proving the bound of interaction blocks. We also thank the anonymous reviewers for helpful comments. This work was supported in part by ARO Grant W911NF-19-1-0302, ARO MURI Grant W911NF-20-1-0082, and NSF Eager Grant 2037687.
\end{acknowledgments}

\bibliography{main}

\clearpage
\newpage
\begin{appendices}
\onecolumngrid
\setcounter{figure}{0}
\setcounter{equation}{0}
\renewcommand{\thefigure}{\Alph{section}\arabic{figure}}
\renewcommand{\theequation}{\Alph{section}\arabic{equation}}
\section{The derivation of the $4-$th moment inequality} \label{appendix:derivation of 4 moment inequality}

The summation of absolute values in \eqref{eqn:magic measure} is tedious to evaluate, so we transform it to the evaluation of the $4-$th moment of Pauli strings as follows. The Pauli strings form a complete basis on $n$ qudits, so an operator $O$ can always be expanded as
\begin{equation}
    O = \sum_{\vec{a}}c_{\vec{a}}P_{\vec{a}}\;,
\end{equation}
where $c_{\vec{a}}$ are the expansion coefficients defined as
\begin{equation}
    c_{\vec{a}} = \frac{1}{d^n}{\rm tr}[OP_{\vec{a}}]\;.
\end{equation}

If we define a $d^n-$dimensional vector $C = \{ c_{\vec{a}}\}$, the magic in \eqref{eqn:magic measure} is the 1-norm $\Vert C\Vert_1$ of $C$ when we take $O = \rho_{\psi} = \ket{\psi}\bra{\psi}$. Then we define a properly normalized vector $V=\{ v_{\vec{a}} \}$ with $v_{\vec{a}} = d^{n/2} c_{\vec{a}}$, whose 2-norm $\Vert V \Vert_{2} = 1$ because
\begin{flalign}
    \Vert V \Vert_{2}^2 = \sum_{\vec{a}}|v_{\vec{a}}|^2 = d^n \sum_{\vec{a}}|c_{\vec{a}}|^2 = {\rm tr}[\rho_{\psi}^2] = \bra{\psi}\ket{\psi}^2 =1\;.
\end{flalign}

The normalized vector $V$ satisfies the following relation between its 1-norm and 4-norm
\begin{equation} \label{eqn:The 4-norm relation}
    \Vert V \Vert_{1} \geq \frac{1}{\Vert V \Vert_{4}^{2}}\;.
\end{equation}

The above inequality is proven as follows: for a normalized vector $V={v_{i}}$, we have the following decomposition
\begin{equation}
    1 = \sum_{i}|v_{i}|^2 = \sum_{i}|b_{i}c_{i}|\;,
\end{equation}
where the second equality is an assumption, and $b_{i},c_{i}$ should be determined later. From the Cauchy-Schwarz inequality, we have
\begin{equation}
    \sum_{i}|b_{i}c_{i}|\leq \left( \sum_{i}|b_{i}|^{3}\right)^{\frac{1}{3}}\left(\sum_{i}|c_{i}|^{\frac{3}{2}}\right)^{\frac{2}{3}}\;.
\end{equation}

Then we assume $b_{i}$ and $c_{i}$ satisfy the following
\begin{equation}
    b_{i} = |v_{i}|^{a}, \quad c_{i} = |v_{i}|^{b}\;,
\end{equation}
thus $a$ and $b$ should satisfy
\begin{equation}
    a+b=2,\quad 3a = 4, \quad \frac{3}{2}b = 1\;,
\end{equation}
where the first equality is from the decomposition, and the second and third ones relate the 1-norm and 4-norm. We can check that they are satisfied, so we have
\begin{equation}
    1\leq \left(\sum_{i}|v_{i}|^4\right)^{\frac{1}{2}} \left(\sum_{i}|v_{i}| \right) = \Vert V \Vert_{4}^{2} \Vert V \Vert_{1}\;.
\end{equation}

By considering the explicit form of $\Vert V\Vert_{4}^2$ as
\begin{equation}
    \Vert V \Vert_{4}^2 
    = d^n (\sum_{\vec{a}} |c_{\vec{a}}|^4)^{1/2} 
    = d^{-n} (\sum_{\vec{a}}{\rm tr}[(P_{\vec{a}}\ket{\psi}\bra{\psi})^{\otimes 4}])^{1/2}\;,
\end{equation}
one can obtain the lower bound of magic in \eqref{eqn:inequality for magic}, which is also equivalent to an inequality relating the $1/2$-Stabilizer R\'enyi Entropy and the $2$-Stabilizer R\'enyi Entropy~\cite{Leone2022}.

\section{The Weingarten calculus and inner products}\label{appendix:The Weingarten calculus and inner products}
In Section \ref{sec:partition functions and interactions}, we define the tensor network at a single site as
\begin{equation}
    \sum_{\{S_{4}\}^2}\begin{tikzpicture}[baseline={([yshift=-.5ex]current bounding box.center)},inner sep=1mm]
    \tikzset{decoration={snake,amplitude=.5mm,segment length=2mm,post length=0mm,pre length=0mm},
    Paulistring/.style = {rectangle, draw=black, thick, fill=blue!25, minimum width=7.2cm,minimum height = 0.6cm},}
    \node [fill=black,circle,inner sep=0pt, minimum size=2mm] (u1) at (0,0) {};
    \node [fill=black,circle,inner sep=0pt, minimum size=2mm] (d1) at (0,-1) {};
    \draw[decorate,thick] (u1) -- (d1);
    \draw[-,draw=blue] (d1) -- +(0.5,0.5);
    \draw[-,draw=blue] (u1) -- +(-0.5,-0.5);
    \draw[-,draw=red] (u1) -- +(0,0.3);
    \draw[-,draw=red] (d1) -- +(0,-0.3);
    \end{tikzpicture}
    = \int_{Haar} dU_{i}
    \begin{tikzpicture}[baseline={([yshift=-.5ex]current bounding box.center)},inner sep=1mm]
    \tikzset{decoration={snake,amplitude=.5mm,segment length=2mm,post length=0mm,pre length=0mm},
    Paulistring/.style = {rectangle, draw=black, thick, fill=blue!25, minimum width=7.2cm,minimum height = 0.6cm},}
    \foreach \i in {8,...,1} {
        \ifnum \i>4
        \node[tensor_blue] (U\i) at (0.4*\i,0.2*\i) {$\overline{U}_{i}$};
        \else  
        \node[tensor_blue] (U\i) at (0.4*\i,0.2*\i) {$U_{i}$};
        \fi
        \draw[-,draw = red] (U\i.north) -- +(0,0.3);
        \draw[-,draw = red] (U\i.south) -- +(0,-0.3);
        \draw[-,draw = blue] (U\i.east) -- +(0.7,0);
        \draw[-,draw = blue] (U\i.west) -- +(-0.7,0);
    };
    \end{tikzpicture},
\end{equation}
where the summation is over all permuatations in $S_{4}$ and the integration is over the Haar measure. The wavy line represents the Weingarten function ${\rm Wg}(\sigma^{-1}\pi,q)$. In our paper, we only require the $t=4$ case, i.e. $\sigma,\pi \in S_{4}$, in which the Weingarten function is listed in Table \ref{tab:S4 Weingarten function}. The ``label" column represents the position of each permutation in the basis $\ket{\sigma}$, namely $\ket{\sigma} =\{ \ket{\mathbb{I}},\ket{(12)},\ket{(13)},\cdots,\ket{(1423)},\ket{(1432)}\}$. In this basis, the 4-th moment operator of Haar random unitaries can be written as a $24\times 24$ Weingarten matrix as given in the supplementary Mathematica notebook.
\begin{table*}
\centering
\begin{tabular}{c|c|c|c}
$s = \sigma^{-1}\pi$ & label & numerator of ${\rm Wg}(s,q)$ & Permutations\\\hline
$\{ 1,1,1,1 \}$ & $1$ & $q^4-8q^2+6$ & $\mathbb{I}$ \\
$\{ 2,1,1 \}$ & $2-7$ & $-q^3+4q$ & $(12),(13),(14),(23),(24),(34)$\\
$\{ 2,2 \}$ & $8-10$ & $q^2+6$ & $((12),(34)),((13),(24)),((14),(23))$ \\
$\{ 3,1 \}$ & $11-18$ & $2q^2-3$ & $(123),(132),(124),(142),(134),(143),(234),(243)$ \\
$\{4 \}$ & $19-24$ & $-5q$ & $(1234),(1243),(1324),(1342),(1423),(1432)$
\end{tabular}
\caption{\label{tab:powers}The list of permutations $s=\sigma^{-1}\pi$ and their corresponding numerator of ${\rm Wg}(s,q)$, with the common denominator $q^2(q^2-1)(q^2-2)(q^2-3)$. The label represents their positions in the basis.}
\label{tab:S4 Weingarten function}
\end{table*}

\begin{table*}
\centering
\begin{tabular}{c|c|c|c|c}
$\sigma $ & label & $\bra{\sigma}\ket{O}$ & Pauli($q=d$)& Permutation list\\\hline
$\{ 1,1,1,1 \}$ & $1$ & $({\rm tr}O)^{4}$& $0$ & $\mathbb{I}$ \\
$\{ 2,1,1 \}$ & $2-7$ & ${\rm tr}O^2({\rm tr}O)^2$& $0$ & $(12),(13),(14),(23),(24),(34)$\\
$\{ 2,2 \}$ & $8-10$ & $({\rm tr}O^2)^2$ & $d^2$ or $0$ & $((12),(34)),((13),(24)),((14),(23))$ \\
$\{ 3,1 \}$ & $11-18$ & ${\rm tr}O^3{\rm tr}O$ & 0 & $(123),(132),(124),(142),(134),(143),(234),(243)$ \\
$\{4 \}$ & $19-24$ & ${\rm tr}O^4$ & $d$ or $0$ & $(1234),(1243),(1324),(1342),(1423),(1432)$
\end{tabular}
\caption{The list of permutations $\sigma\in S_{4}$ and their corresponding inner product $\bra{\sigma}\ket{O}$. The label represents their positions in the basis. The fourth column summarizes the contractions for $O$ being single Pauli operators with local dimension $q=d$}
\label{tab:S4 operator inner product}
\end{table*}

\begin{table*}
\centering
\begin{tabular}{c|c|c|c}
$s = \sigma^{-1}\pi$ & label & Inner product of $\bra{\sigma}\ket{\pi}$ & Permutation list\\\hline
$\{ 1,1,1,1 \}$ & $1$ & $q^4$ & $\mathbb{I}$ \\
$\{ 2,1,1 \}$ & $2-7$ & $q^3$ & $(12),(13),(14),(23),(24),(34)$\\
$\{ 2,2 \}$ & $8-10$ & $q^2$ & $((12),(34)),((13),(24)),((14),(23))$ \\
$\{ 3,1 \}$ & $11-18$ & $q^2$ & $(123),(132),(124),(142),(134),(143),(234),(243)$ \\
$\{4 \}$ & $19-24$ & $q$ & $(1234),(1243),(1324),(1342),(1423),(1432)$
\end{tabular}
\caption{The contractions $\bra{\sigma}\ket{\pi}$ for $\sigma,\pi\in S_{4}$ and their corresponding $s=\sigma^{-1}\pi$. The label represents their positions in the basis.}
\label{tab:S4 inner product}
\end{table*}
In Eq.\ref{eqn:expectation of Pa local operators}, the local operator is defined as
\begin{equation}
    \begin{tikzpicture}[baseline={([yshift=-.5ex]current bounding box.center)},inner sep=1mm]
    \node[tensor_blue] (O) at (0,0) {$I \otimes O_1$}; 
    \draw[-,draw=red] (O.north) -- +(0,0.3);
    \draw[-,draw=red] (O.south) -- +(0,-0.3);
    \end{tikzpicture}
    \equiv 
    \begin{tikzpicture}[baseline={([yshift=-.5ex]current bounding box.center)},inner sep=1mm]
    \foreach \i in {1,...,4} {
    \draw[-,draw=red] (0.4*\i,0.55) -- (0.4*\i,-0.55);
    };
    \foreach \i in {5,...,8} {
    \node[tensor_blue] (O\i) at (0.7*\i-1.4,0) {$O_{1}$};
    \draw[-,draw=red] (O\i.north) -- +(0,0.3);
    \draw[-,draw=red] (O\i.south) -- +(0,-0.3);
};
    \end{tikzpicture}\ ,
\end{equation}
where on the right hand side, the vertical red lines are the identity operator, and the $O_{1}$ blocks are the $O_{1}$ operators on one qudit. The first and last four copies are in the $\ket{\psi}^{\otimes4}$ and $\overline{\ket{\psi}}^{\otimes 4}$ subspaces, respectively. In this convention, the maximally entangled state $\ket{\Omega}$ is
\begin{equation}
    \ket{\Omega} = 
    \begin{tikzpicture}[baseline={([yshift=-.5ex]current bounding box.center)},inner sep=1mm]
    \foreach \i in {1,...,4}{
    \coordinate (l\i) at (0.3*\i,0);
    \coordinate (r\i) at (1.4+0.3*\i,0);
    \draw [red]   (l\i) to[out=-80,in=260] (r\i);
    \node [fill=red,circle,inner sep=0pt, minimum size=2mm,label=above:$\i$] at (l\i) {};
    \node [fill=red,circle,inner sep=0pt, minimum size=2mm,label=above:$\i$] at (r\i) {};
    }
    \end{tikzpicture}.
\end{equation}
Thus the state $\ket{O}$ is
\begin{equation}
    \ket{O} = 
    \begin{tikzpicture}[baseline={([yshift=-.5ex]current bounding box.center)},inner sep=1mm]
    \foreach \i in {1,...,4} {
    \coordinate (d\i) at (0.4*\i,-0.55);
    \draw[-,draw=red] (0.4*\i,0.55) -- (d\i);
    };
    \foreach \i in {5,...,8} {
    \coordinate (d\i) at (0.7*\i-1.4,-0.55);
    \node[tensor_blue] (O\i) at (0.7*\i-1.4,0) {$O_{1}$};
    \draw[-,draw=red] (O\i.north) -- +(0,0.3);
    \draw[-,draw=red] (O\i.south) -- (d\i);
    };
    \foreach \i in {1,...,4} {
        \pgfmathtruncatemacro{\iplusfour}{\i + 4};
        \draw[red] (d\i) ..controls(0.55*\i+0.15,-0.8) and (0.55*\i+1.25,-0.8) .. (d\iplusfour);
    };
    \end{tikzpicture}\ .
\end{equation}
For convenience, we now draw the diagram vertically and use different colors to represent the four legs. The left and right four legs in the original diagram are moved to the upper and bottom four legs in the new diagram, respectively. Therefore, some of the states are
\begin{equation}
    \ket{\Omega} =
    \begin{tikzpicture}[baseline={([yshift=-.5ex]current bounding box.center)},inner sep=1mm]
    \coordinate (1s) at (0,0);
    \coordinate (2s) at (0,-0.3);
    \coordinate (3s) at (0,-0.6);
    \coordinate (4s) at (0,-0.9);
    \coordinate (1e) at (0,-1.4);
    \coordinate (2e) at (0,-1.7);
    \coordinate (3e) at (0,-2.0);
    \coordinate (4e) at (0,-2.3);

    \draw [black]   (1s) to[out=-10,in=10] (1e);
    \draw [red]   (2s) to[out=-10,in=10] (2e);
    \draw [evergreen] (3s) to[out=-10,in=10] (3e);
    \draw [blue] (4s) to[out=-10,in=10] (4e);
    
    \node [fill=black,circle,inner sep=0pt, minimum size=2mm,label=left:$1$] at (1s) {};
    \node [fill=red,circle,inner sep=0pt, minimum size=2mm,label=left:$2$] at (2s) {};
    \node [fill=evergreen,circle,inner sep=0pt, minimum size=2mm,label=left:$3$] at (3s) {};
    \node [fill=blue,circle,inner sep=0pt, minimum size=2mm,label=left:$4$] at (4s) {};
    \node [fill=black,circle,inner sep=0pt, minimum size=2mm,label=left:$1$] at (1e) {};
    \node [fill=red,circle,inner sep=0pt, minimum size=2mm,label=left:$2$] at (2e) {};
    \node [fill=evergreen,circle,inner sep=0pt, minimum size=2mm,label=left:$3$] at (3e) {};
    \node [fill=blue,circle,inner sep=0pt, minimum size=2mm,label=left:$4$] at (4e) {};
    \end{tikzpicture},
    \ket{(12)} = 
    \begin{tikzpicture}[baseline={([yshift=-.5ex]current bounding box.center)},inner sep=1mm]
    \coordinate (1s) at (0,0);
    \coordinate (2s) at (0,-0.3);
    \coordinate (3s) at (0,-0.6);
    \coordinate (4s) at (0,-0.9);
    \coordinate (2e) at (0,-1.4);
    \coordinate (1e) at (0,-1.7);
    \coordinate (3e) at (0,-2.0);
    \coordinate (4e) at (0,-2.3);

    \draw [black]   (1s) to[out=-10,in=10] (1e);
    \draw [red]   (2s) to[out=-10,in=10] (2e);
    \draw [evergreen] (3s) to[out=-10,in=10] (3e);
    \draw [blue] (4s) to[out=-10,in=10] (4e);
    
    \node [fill=black,circle,inner sep=0pt, minimum size=2mm,label=left:$1$] at (1s) {};
    \node [fill=red,circle,inner sep=0pt, minimum size=2mm,label=left:$2$] at (2s) {};
    \node [fill=evergreen,circle,inner sep=0pt, minimum size=2mm,label=left:$3$] at (3s) {};
    \node [fill=blue,circle,inner sep=0pt, minimum size=2mm,label=left:$4$] at (4s) {};
    \node [fill=black,circle,inner sep=0pt, minimum size=2mm,label=left:$1$] at (1e) {};
    \node [fill=red,circle,inner sep=0pt, minimum size=2mm,label=left:$2$] at (2e) {};
    \node [fill=evergreen,circle,inner sep=0pt, minimum size=2mm,label=left:$3$] at (3e) {};
    \node [fill=blue,circle,inner sep=0pt, minimum size=2mm,label=left:$4$] at (4e) {};
    \end{tikzpicture},
    \bra{(123)} = 
    \begin{tikzpicture}[baseline={([yshift=-.5ex]current bounding box.center)},inner sep=1mm]
    \coordinate (1s) at (0,0);
    \coordinate (2s) at (0,-0.3);
    \coordinate (3s) at (0,-0.6);
    \coordinate (4s) at (0,-0.9);
    \coordinate (3e) at (0,-1.4);
    \coordinate (1e) at (0,-1.7);
    \coordinate (2e) at (0,-2.0);
    \coordinate (4e) at (0,-2.3);

    \draw [black]   (1s) to[out=-170,in=170] (1e);
    \draw [red]   (2s) to[out=-170,in=170] (2e);
    \draw [evergreen] (3s) to[out=-170,in=170] (3e);
    \draw [blue] (4s) to[out=-170,in=170] (4e);
    
    \node [fill=black,circle,inner sep=0pt, minimum size=2mm,label=right:$1$] at (1s) {};
    \node [fill=red,circle,inner sep=0pt, minimum size=2mm,label=right:$2$] at (2s) {};
    \node [fill=evergreen,circle,inner sep=0pt, minimum size=2mm,label=right:$3$] at (3s) {};
    \node [fill=blue,circle,inner sep=0pt, minimum size=2mm,label=right:$4$] at (4s) {};
    \node [fill=black,circle,inner sep=0pt, minimum size=2mm,label=right:$1$] at (1e) {};
    \node [fill=red,circle,inner sep=0pt, minimum size=2mm,label=right:$2$] at (2e) {};
    \node [fill=evergreen,circle,inner sep=0pt, minimum size=2mm,label=right:$3$] at (3e) {};
    \node [fill=blue,circle,inner sep=0pt, minimum size=2mm,label=right:$4$] at (4e) {};
    \end{tikzpicture},
\end{equation}
where the permutation state $\ket{\sigma} = (\mathbb{I}\otimes r(\sigma))\ket{\Omega}$ is obtained by permuting the bottom legs according to $\sigma$. We can calculate the inner product as (taking the $\bra{((12),(34))}\ket{O}$ as an example)
\begin{flalign}
   &\bra{((12),(34))}\ket{O} =\notag\\
   &\begin{tikzpicture}[baseline={([yshift=-.5ex]current bounding box.center)},inner sep=1mm]
    \coordinate (1s) at (0,0);
    \coordinate (2s) at (0,-0.3);
    \coordinate (3s) at (0,-0.6);
    \coordinate (4s) at (0,-0.9);
    \coordinate (2e) at (0,-2.0);
    \coordinate (1e) at (0,-2.5);
    \coordinate (4e) at (0,-3.0);
    \coordinate (3e) at (0,-3.5);
    \draw [black]   (1s) to[out=-170,in=170] (1e);
    \draw [red]   (2s) to[out=-170,in=170] (2e);
    \draw [evergreen] (3s) to[out=-170,in=170] (3e);
    \draw [blue] (4s) to[out=-170,in=170] (4e);
    \node [fill=black,circle,inner sep=0pt, minimum size=2mm] at (1s) {};
    \node [fill=red,circle,inner sep=0pt, minimum size=2mm] at (2s) {};
    \node [fill=evergreen,circle,inner sep=0pt, minimum size=2mm] at (3s) {};
    \node [fill=blue,circle,inner sep=0pt, minimum size=2mm] at (4s) {};
    \node [fill=black,circle,inner sep=0pt, minimum size=2mm] at (1e) {};
    \node [fill=red,circle,inner sep=0pt, minimum size=2mm] at (2e) {};
    \node [fill=evergreen,circle,inner sep=0pt, minimum size=2mm] at (3e) {};
    \node [fill=blue,circle,inner sep=0pt, minimum size=2mm] at (4e) {};
    \coordinate (1s1) at (0.5,0);
    \coordinate (2s1) at (0.5,-0.3);
    \coordinate (3s1) at (0.5,-0.6);
    \coordinate (4s1) at (0.5,-0.9);
    \coordinate (1e1) at (0.5,-2.0);
    \coordinate (2e1) at (0.5,-2.5);
    \coordinate (3e1) at (0.5,-3.0);
    \coordinate (4e1) at (0.5,-3.5);
    \draw [black]   (1s1) to[out=-10,in=0] (1e1.east);
    \draw [red]   (2s1) to[out=-10,in=0] (2e1.east);
    \draw [evergreen] (3s1) to[out=-10,in=0] (3e1.east);
    \draw [blue] (4s1) to[out=-10,in=0] (4e1.east);
    \node [fill=black,circle,inner sep=0pt, minimum size=2mm] at (1s1) {};
    \node [fill=red,circle,inner sep=0pt, minimum size=2mm] at (2s1) {};
    \node [fill=evergreen,circle,inner sep=0pt, minimum size=2mm] at (3s1) {};
    \node [fill=blue,circle,inner sep=0pt, minimum size=2mm] at (4s1) {};
    \node [tensor_blue] at (1e1) {$O$};
    \node [tensor_blue] at (2e1) {$O$};
    \node [tensor_blue] at (3e1) {$O$};
    \node [tensor_blue] at (4e1) {$O$};
\end{tikzpicture}=
\begin{tikzpicture}[baseline={([yshift=-.5ex]current bounding box.center)},inner sep=1mm]
    \node[tensor_blue] (O1) at (0,0) {$O$};
    \node[tensor_blue] (O2) at (1,0) {$O$};
    \draw[-,red] (O1.east)--(O2.west);
    \draw[-,black] (O1.west) .. controls +(-1, 1) and +(1, 1) .. (O2.east);
    \node[tensor_blue] (O3) at (0,-1.5) {$O$};
    \node[tensor_blue] (O4) at (1,-1.5) {$O$};
    \draw[-,blue] (O3.east)--(O4.west);
    \draw[-,evergreen] (O3.west) .. controls +(-1, 1) and +(1, 1) .. (O4.east);
\end{tikzpicture} 
= ({\rm tr}O^2)^2.
\end{flalign}
All other inner products $\bra{\sigma}\ket{O}$ can be calculated similarly, and we summarize them in Table \ref{tab:S4 operator inner product}, where in the fourth column we list the results for $O$ being single Pauli operators with local dimension $d$. Since Pauli operators are traceless, the 1st, 2nd and 4th rows are $0$. The 3rd and 5th rows are nonzero only when $O^2=I$ or $O^4=I$, respectively, which are a subset of the local Pauli operators, defined as $O_{1}$ and $O_{2}$ in Section \ref{sec:partition functions and interactions}.

The inner product of permutation states can be calculated as (taking the $\bra{(1423)}\ket{(123)}$ as an example)
\begin{equation}
    \bra{(1423)}\ket{(123)}=
\begin{tikzpicture}[baseline={([yshift=-.5ex]current bounding box.center)},inner sep=1mm]
    \coordinate (1s) at (0,0);
    \coordinate (2s) at (0,-0.3);
    \coordinate (3s) at (0,-0.6);
    \coordinate (4s) at (0,-0.9);
    \coordinate (3e) at (0,-1.4);
    \coordinate (4e) at (0,-1.7);
    \coordinate (2e) at (0,-2.0);
    \coordinate (1e) at (0,-2.3);

    \draw [black]   (1s) to[out=-170,in=170] (1e);
    \draw [red]   (2s) to[out=-170,in=170] (2e);
    \draw [evergreen] (3s) to[out=-170,in=170] (3e);
    \draw [blue] (4s) to[out=-170,in=170] (4e);
    
    \node [fill=black,circle,inner sep=0pt, minimum size=2mm] at (1s) {};
    \node [fill=red,circle,inner sep=0pt, minimum size=2mm] at (2s) {};
    \node [fill=evergreen,circle,inner sep=0pt, minimum size=2mm] at (3s) {};
    \node [fill=blue,circle,inner sep=0pt, minimum size=2mm] at (4s) {};
    \node [fill=black,circle,inner sep=0pt, minimum size=2mm] at (1e) {};
    \node [fill=red,circle,inner sep=0pt, minimum size=2mm] at (2e) {};
    \node [fill=evergreen,circle,inner sep=0pt, minimum size=2mm] at (3e) {};
    \node [fill=blue,circle,inner sep=0pt, minimum size=2mm] at (4e) {};
\end{tikzpicture}
\ 
\begin{tikzpicture}[baseline={([yshift=-.5ex]current bounding box.center)},inner sep=1mm]
    \coordinate (1s) at (0,0);
    \coordinate (2s) at (0,-0.3);
    \coordinate (3s) at (0,-0.6);
    \coordinate (4s) at (0,-0.9);
    \coordinate (3e) at (0,-1.4);
    \coordinate (1e) at (0,-1.7);
    \coordinate (2e) at (0,-2.0);
    \coordinate (4e) at (0,-2.3);

    \draw [black]   (1s) to[out=-10,in=10] (1e);
    \draw [red]   (2s) to[out=-10,in=10] (2e);
    \draw [evergreen] (3s) to[out=-10,in=10] (3e);
    \draw [blue] (4s) to[out=-10,in=10] (4e);
    
    \node [fill=black,circle,inner sep=0pt, minimum size=2mm] at (1s) {};
    \node [fill=red,circle,inner sep=0pt, minimum size=2mm] at (2s) {};
    \node [fill=evergreen,circle,inner sep=0pt, minimum size=2mm] at (3s) {};
    \node [fill=blue,circle,inner sep=0pt, minimum size=2mm] at (4s) {};
    \node [fill=black,circle,inner sep=0pt, minimum size=2mm] at (1e) {};
    \node [fill=red,circle,inner sep=0pt, minimum size=2mm] at (2e) {};
    \node [fill=evergreen,circle,inner sep=0pt, minimum size=2mm] at (3e) {};
    \node [fill=blue,circle,inner sep=0pt, minimum size=2mm] at (4e) {};
    \end{tikzpicture}
    =
    \begin{tikzpicture}[baseline={([yshift=-.5ex]current bounding box.center)},inner sep=1mm,
    dot/.style = {circle, draw, minimum size=0.6 cm,
              inner sep=0pt, outer sep=0pt}]
    \node[dot,black] (1) at (0,0) {};
    \node[dot,red] (1) at (0,-0.8) {};
    \node[dot,evergreen] (1) at (0,-1.6) {};
    \end{tikzpicture}
    =q^3.
\end{equation}
Therefore, calculating $\bra{\sigma}\ket{\pi}$ is simply getting $s=\sigma^{-1}\pi$ and checking the number of closed permutations (loops) in $s$, and we summarize the results in Table \ref{tab:S4 inner product}.

\section{More analysis on the upper bound}\label{appendix:more analysis on the upper bound}
The explicit calculation of the eigenvalues for the three blocks is in the supplementary Mathematica notebook. In the notebook, we check the eigenvalues of the three blocks are non-negative when $d\geq 2, B\geq 2$. We also thoroughly check \eqref{eqn: The spectral radii} and \eqref{eqn: The spectral radii at d=2} for each eigenvalue. All above statements are also checked explicitly by the method described in this section in supplementary materials. Here we provide a simplified version of the proof. A useful fact is, when $d\geq 2, B\geq 2$, the common denominator $B^2d^2(B^6d^6-6B^4d^4+11B^2d^2-6)$ of the eigenvalues (from the common denominator of the Weingarten function) is positive because
\begin{flalign}\label{eqn:denominator inequality}
    B^6d^6-6B^4d^4+11B^2d^2-6 &\geq 10B^4 d^4 + 11B^2d^2 -6 \notag\\
    &\geq 1611B^2d^2-6>0\;,
\end{flalign}
where we have used $B\geq 2,d\geq 2$ to transform the higher order terms in $B,d$ to lower order terms and collect the terms at the same order. This trick will be applied repeatedly in the following sections.
\subsection{The green block}
As shown in the supplementary Mathematica notebook, the spectral radius of the green block is
\begin{equation}
    \rho_1  \equiv\rho(\begin{tikzpicture}[baseline={([yshift=-.5ex]current bounding box.center)},inner sep=0mm]
        \node[tensor_green] (O1) at (-4.8,0) {};
        \node[inner sep=0pt, minimum size=2mm] (O1left) at (-5.4,0) {};
        \node[inner sep=0pt, minimum size=2mm] (O1right) at (-4.2,0) {};
        \draw[-](O1) -- (O1left);
        \draw[-](O1) -- (O1right);
        \end{tikzpicture}) = \frac{B^4 d^4-13 B^2 d^2+36}{\left(B^2 d^2-3\right) \left(B^2 d^2-2\right)}\;.
\end{equation}
To prove $\rho_{1}\leq 1$, we can equivalently prove
\begin{equation}
    8 B^2 d^2-30\geq 0\;,
\end{equation}
which is satisfied when $B\geq 2,d\geq 2$, so $\rho_{1}\leq 1$.
\subsection{The blue block}
As shown in the supplementary Mathematica notebook, the spectral radius of the blue block is
\begin{equation}
        \rho_{2} \equiv \rho(\begin{tikzpicture}[baseline={([yshift=-.5ex]current bounding box.center)},inner sep=0mm]
        \node[tensor_blue] (O1) at (-4.8,0) {};
        \node[inner sep=0pt, minimum size=2mm] (O1left) at (-5.4,0) {};
        \node[inner sep=0pt, minimum size=2mm] (O1right) at (-4.2,0) {};
        \draw[-](O1) -- (O1left);
        \draw[-](O1) -- (O1right);
        \end{tikzpicture}) = \frac{A_{1}+(d-1)\sqrt{A_{2}}}{A_{3}}\;,
\end{equation}
where the $A_{1},A_{2}$ and $A_{3}$ are polynomials in terms of $d,B$, which are
    \begin{flalign}
        A_{1} = &B^6 \left(d^4+d^3\right)+B^4 \left(2 d^4-5 d^3-26 d^2+d\right)+B^2 \left(-8 d^3+23
   d^2+65 d+18\right)-18 d-54\;,\notag\\
   A_{2} = &B^{12} d^6+B^{10} \left(4 d^6+6 d^5+2 d^4\right)+B^8 \left(4 d^6-36 d^5-213 d^4-158
   d^3+d^2\right)\notag\\
   &+B^6 \left(-24 d^5+156 d^4+1078 d^3+1282 d^2+36 d\right)+B^4 \left(64
   d^4-488 d^3-2075 d^2-2736 d+324\right)\notag\\
   &+B^2 \left(288 d^2+1836 d+648\right)+324\;,\notag\\
    A_3 = &2 \left(B^6 d^6-6 B^4 d^4+11 B^2 d^2-6\right)\;.
\end{flalign}
Since $A_{3}$ is positive, $\rho_{2}\leq 2/d^2$ is equivalent to
\begin{equation} \label{eqn: Inequality rho2}
    d^2(d-1)\sqrt{A_{2}}\leq 2A_{3}-d^2A_{1}\;.
\end{equation}
One can check both sides are positive. As an example, we show $A_{2}\geq 0$ for $d,B\geq 0$. We firstly check when $B\geq 3, d\geq 2$
    \begin{flalign}
        B^{12} d^6 &\geq 9B^{10}d^{6}\\
        B^{10}\left(4 d^6+6 d^5+2 d^4\right) &\geq B^{10}\left(4 d^6+6 d^5+2 d^4\right)\\
        B^8 \left(4 d^6-36 d^5-213 d^4-158
   d^3+d^2\right) &\geq B^8 \left(4 d^6-36 d^5-213 d^4-158d^3\right) \\
   B^6 \left(-24 d^5+156 d^4+1078 d^3+1282 d^2+36 d\right) &\geq B^6 \left(-24 d^5+156 d^4\right) \\
   B^4 \left(64
   d^4-488 d^3-2075 d^2-2736 d+324\right) &\geq B^4 \left(64
   d^4-488 d^3-3443 d^2\right) \\
   B^2 \left(288 d^2+1836 d+648\right)+324 &\geq 0
    \end{flalign}

Therefore, when $B\geq 3,d\geq 2$, $A_{2}$ satisfies
\begin{flalign}
    A_{2}|_{B\geq 3,d\geq 2}&\geq  B^8 \left(121 d^6+18 d^5-195 d^4-158d^3\right)+B^6 \left(-24 d^5+156 d^4\right)+B^4 \left(64d^4-488 d^3-3443 d^2\right) \notag\\
   &\geq 1508B^8d^3 + 3B^6d^5 + 1453B^4d^2\geq 0\;.
\end{flalign}    

Since $d$ and $B$ are integers, now we only need to focus on $d\geq 2,B = 2$ case, which gives us
\begin{flalign}
    A_{2}|_{B=2} &= 9216 d^6-4608 d^5-41472 d^4\notag\\
    &+20736 d^3+50256 d^2-34128 d+8100\;.
\end{flalign}
When $d=2$, we have 
\begin{equation}
    A_{2}|_{B=2,d=2} = 85572 \geq 0
\end{equation}
when $d\geq 3$, we have
\begin{flalign}
    A_{2}|_{B=2,d\geq 3} \geq 27648d^4 + 116640d \geq 0\;.
\end{flalign}
So we complete the proof $A_{2}\geq 0$ for $d\geq 2, B\geq 2$. We have checked all other inequality like $2A_{3}-d^2A_{2}$ can be proved by this similar method. Those polynomials always have a highest order term in $B,d$ with positive coefficient, so when $d$ and $B$ are large, for example $d\geq k, B\geq l$ for some integers $k,l \geq 2$, we can always use the scaling method to prove their positivity. Then we only need to check those polynomials are non-negative for some limited cases when $2\leq d\leq k$ and $2\leq B \leq l$. So we can square both sides in \eqref{eqn: Inequality rho2} to transform the inequality to
    \begin{flalign}
        &B^{12} \left(8 d^{12}-4 d^{11}\right)+B^{10} \left(-16 d^{12}+36 d^{11}+8 d^{10}-12
   d^9\right)+B^8 \left(72 d^{11}-552 d^9+176 d^8+104 d^7\right)\notag\\&+B^6 \left(-8 d^{11}-32
   d^{10}-572 d^9+144 d^8+1388 d^7-1120 d^6-40 d^5\right)\notag\\&+B^4 \left(176 d^9+488
   d^8+3092 d^7-416 d^6-344 d^5+256 d^4+48 d^3\right)\notag\\&+B^2 \left(-1224 d^7-2088 d^6-7008
   d^5+3264 d^4+3120 d^3-1248 d^2\right)+2592 d^5+2592 d^4-864 d^3-2592 d^2+576\geq 0
    \end{flalign}
By appropriately using $d\geq 2$, one can obatin an lower bound on each term as
\begin{flalign}
    B^{12} \left(8 d^{12}-4 d^{11}\right) &\geq B^{12}d^6\;, \notag\\
    B^{10} \left(-16 d^{12}+36 d^{11}+8 d^{10}-12
   d^9\right) &\geq B^{10}(-16d^{12}+36d^{10})\;,\notag\\
   B^8 \left(72 d^{11}-552 d^9+176 d^8+104 d^7\right) \geq -264B^{8}d^9\;,\notag\\
    B^6 \left(-8 d^{11}-32
   d^{10}-572 d^9+144 d^8+1388 d^7-1120 d^6-40 d^5\right)&\geq B^{6}(-167d^{11}+144d^8)\;,\notag\\
    B^4 \left(176 d^9+488
   d^8+3092 d^7-416 d^6-344 d^5+256 d^4+48 d^3\right)&\geq 176B^4d^9 \;,\notag\\
       B^2 \left(-1224 d^7-2088 d^6-7008
   d^5+3264 d^4+3120 d^3-1248 d^2\right)&\geq -4020B^2d^7\;,\notag\\
    2592 d^5+2592 d^4-864 d^3-2592 d^2+576&\geq 0\;.
\end{flalign} 
So we only need to prove
\begin{equation}
    B^{12}d^6-16B^{10}d^{12}+36B^{10}d^{10}-264B^8d^9-167B^6d^{11}+144B^6d^{8}+176B^4d^9-4020B^2d^7 \geq 0\;.
\end{equation}
By the same trick, we can prove
\balance
\begin{equation}
    {\rm LHS} \geq 20B^8d^9+89B^6d^{11}+3414B^2d^7\;, 
\end{equation}
where $20B^8d^9+89B^6d^{11}+3414B^2d^7\geq 0$, so we complete the proof, and we conclude that $\rho_{2}\leq 2/d^2$ when $d\geq 2,B\geq 2$.
\subsection{The blue block at $d=2$}
When $d=2$, as given in \eqref{eqn: The spectral radii at d=2}, we have a better bound 
\begin{equation}
    \rho_{2}^{\prime} = \frac{A_{1}+(d-1)\sqrt{A_{2}}}{A_{3}}\bigg\rvert_{d=2} \leq 1/d^2 = 1/4\;,
\end{equation}
which is equivalent to prove 
\begin{equation} \label{eqn: Inequality rho2 d=2}
    4\sqrt{A_{2}}\leq A_{3}-4A_{2}\;.
\end{equation}
or
\begin{flalign}
    &8192 B^{10}+111104 B^8-532096 B^6+813312 B^4\notag\\&-516288 B^2+115920 \geq 0\;.
\end{flalign}
By appropriate scaling using $B\geq 2$, one can prove the LHS of the above inequality satisfies
\begin{equation}
    {\rm LHS} \geq 43392B^6+2736960B^2 \geq 0\;,
\end{equation}
which completes the proof.
\subsection{The purple block}
As shown in the Supplementary notebook, the spectral radius of the purple block is
\begin{flalign}
    \rho_3  \equiv\rho(\begin{tikzpicture}[baseline={([yshift=-.5ex]current bounding box.center)},inner sep=0mm]
        \node[tensor_purple] (O1) at (-4.8,0) {};
        \node[inner sep=0pt, minimum size=2mm] (O1left) at (-5.4,0) {};
        \node[inner sep=0pt, minimum size=2mm] (O1right) at (-4.2,0) {};
        \draw[-](O1) -- (O1left);
        \draw[-](O1) -- (O1right);
        \end{tikzpicture}) = \frac{B^6 d^3+5 B^4 d^3-20 B^4 d^2+B^4 d-16 B^2 d^2+65 B^2 d-36}{\left(B^2 d^2-3\right)
   \left(B^2 d^2-2\right) \left(B^2 d^2-1\right)}\;.    
\end{flalign}    
To prove $\rho_{3}\leq 3/d^3$, we can equivalently prove
    \begin{equation}
        2 B^6 d^6+B^4 \left(-5 d^6+20 d^5-19 d^4\right)+B^2 \left(16 d^5-65 d^4+33
   d^2\right)+36 d^3-18 \geq 0\;.
    \end{equation}
By the scaling method, one can show the LHS of the above inequality satisfies
\begin{flalign}
    {\rm LHS} &\geq 2B^6d^6 -5B^4d^6+16B^2d^5-65B^2d^4\notag\\
    &\geq 3B^4d^6-33B^2d^4\geq 15B^2d^4 \geq 0\;,
\end{flalign}
which completes the proof of $\rho_{3}\leq 3/d^3$. So far we have proved the inequalities in the main text \eqref{eqn: The spectral radii} and \eqref{eqn: The spectral radii at d=2}.

\clearpage
\newpage
\section{More on numerical calculation}\label{appendix:more on numerical calculation}
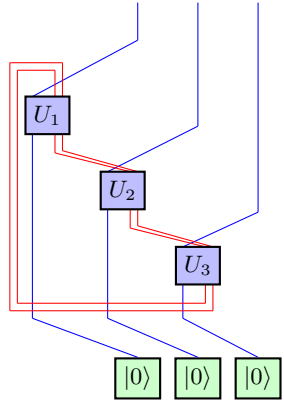
\begin{figure}
\centering
\begin{tikzpicture}
\node[tensor_green] (01) at (1.2,-3.5) {$\ket{0}$};
\node[tensor_green] (02) at (2,-3.5) {$\ket{0}$};
\node[tensor_green] (03) at (2.8,-3.5) {$\ket{0}$};
\coordinate (T1blue) at (-0.2,0.25);
\coordinate (T2blue) at (0.8,-0.75);
\coordinate (T3blue) at (1.8,-1.75);
\coordinate (T1turndown) at (-0.2,-2.7);
\coordinate (T2turndown) at (0.8,-2.7);
\coordinate (T3turndown) at (1.8,-2.7);
\coordinate (T1turnup) at (1.2,1);
\coordinate (T2turnup) at (2,-0.144);
\coordinate (T3turnup) at (2.8,-1.288);
\coordinate (T1up) at (1.2,1.5);
\coordinate (T2up) at (2,1.5);
\coordinate (T3up) at (2.8,1.5);
\coordinate (T1red1) at (0.1,0.25);
\coordinate (T1red2) at (0.2,0.25);
\coordinate (T2red1) at (1.1,-0.75);
\coordinate (T2red2) at (1.2,-0.75);
\coordinate (T3red1) at (2.1,-1.75);
\coordinate (T3red2) at (2.2,-1.75);
\coordinate (T1red1turndown) at (0.1,-0.5);
\coordinate (T1red2turndown) at (0.2,-0.47);
\coordinate (T2red1turndown) at (1.1,-1.5);
\coordinate (T2red2turndown) at (1.2,-1.47);
\coordinate (T3red1turndown) at (2.1,-2.5);
\coordinate (T3red2turndown) at (2.2,-2.6);
\coordinate (red1southwest) at (-0.4,-2.5);
\coordinate (red2southwest) at (-0.5,-2.6);
\coordinate (red1northwest) at (-0.4,0.6);
\coordinate (red2northwest) at (-0.5,0.7);
\coordinate (red1northend) at (0.1,0.6);
\coordinate (red2northend) at (0.2,0.7);
\draw[blue] (T1blue) -- (T1turndown);
\draw[blue] (T2blue) -- (T2turndown);
\draw[blue] (T3blue) -- (T3turndown);
\draw[blue] (T1blue) -- (T1turnup);
\draw[blue] (T2blue) -- (T2turnup);
\draw[blue] (T3blue) -- (T3turnup);
\draw[blue] (T1turnup) -- (T1up);
\draw[blue] (T2turnup) -- (T2up);
\draw[blue] (T3turnup) -- (T3up);
\draw[red] (T1red1) -- (T1red1turndown);
\draw[red] (T1red2) -- (T1red2turndown);
\draw[red] (T2red1) -- (T2red1turndown);
\draw[red] (T2red2) -- (T2red2turndown);
\draw[red] (T3red1) -- (T3red1turndown);
\draw[red] (T3red2) -- (T3red2turndown);
\draw[red] (T1red1turndown) -- (T2red1);
\draw[red] (T1red2turndown) -- (T2red2);
\draw[red] (T2red1turndown) -- (T3red1);
\draw[red] (T2red2turndown) -- (T3red2);
\draw[red] (red1southwest) -- (T3red1turndown);
\draw[red] (red2southwest) -- (T3red2turndown);
\draw[red] (red1southwest) -- (red1northwest);
\draw[red] (red2southwest) -- (red2northwest);
\draw[red] (red1northend) -- (red1northwest);
\draw[red] (red2northend) -- (red2northwest);
\draw[red] (red1northend) -- (T1red1);
\draw[red] (red2northend) -- (T1red2);
\node[tensor_blue] (T1) at (0,0) {$U_{1}$};
\node[tensor_blue] (T2) at (1,-1) {$U_{2}$};
\node[tensor_blue] (T3) at (2,-2) {$U_{3}$};
\draw[blue] (T1turndown) -- (01.north);
\draw[blue] (T2turndown) -- (02.north);
\draw[blue] (T3turndown) -- (03.north);
\end{tikzpicture}
\caption{\label{fig:the contraction of RMPS}The contraction of random unitaries in the RMPS for $n=3, B=4$, where the red lines are the auxiliary qubits being traced out, and the blue lines are the qubits in our RMPS. The $U_{1},U_{2},U_{3}$ are Haar random unitaries.}
\end{figure}

In constructing an RMPS, the contraction over the bonds with dimension $B$ in ~\eqref{eqn:RMPS} is performed by introducing $n_B$ extra auxiliary qubits on the bonds, so the bond dimension is $B=2^{n_B}$. In Figure~\ref{fig:the contraction of RMPS}, we choose $n_B=2$ (the red lines) as an example, where the number $n$ of qubits is $3$ (the blue lines). If we read from the bottom to the top, Figure~\ref{fig:the contraction of RMPS} is a quantum circuit with $5$ qubits, which is composed of swap gates and three random unitaries $U_{1},U_{2},U_{3}$. So one can realize those quantum gates in the simulation, and trace out the two auxiliary qubits, to finally generate the RMPS. This procedure can be generalized to general $n$ and $B=2^{n_B}$ by correspondingly modifying the quantum circuit.

In our simulation, we choose $n_B=1\sim 3$ corresponding to $B=2,4,8$. For each $n$ in $n=2\sim 8$, we generate $100$ RMPSs and calculate the magic of them as in \eqref{eqn:magic measure}, and then take the average over the magic to get the expectation value $\mathbb{E}\ M(\ket{\psi})$, since the unitaries are drawn from Haar random (uniform) measure. By taking the logarithm of $\mathbb{E}\ M(\ket{\psi})$ and plotting them versus the qubit number $n$, we obtain the result in Figure~\ref{fig:The log magic in terms of n small B}.


\clearpage

\end{appendices}

\pagebreak
\widetext
\begin{center}
\textbf{\large Supplementary Materials: Magic of Random Matrix Product States}
\end{center}
\setcounter{section}{0}
\setcounter{subsection}{0}
\setcounter{equation}{0}
\setcounter{figure}{0}
\setcounter{table}{0}
\setcounter{page}{1}
\makeatletter
\renewcommand{\thesection}{S\arabic{section}}
\renewcommand{\theequation}{S\arabic{equation}}
\renewcommand{\thefigure}{S\arabic{figure}}
\renewcommand{\bibnumfmt}[1]{[S#1]}
\renewcommand{\citenumfont}[1]{S#1}

\small
In the main text, we do the calculation for the largest eigenvalue of each interaction block (see the supplementary Mathematica notebook), which serves as the starting point of our simplified proof of the upper bounds. To be rigorous, we should check each eigenvalue of the interaction blocks satisfies the corresponding upper bounds, which will be verified explicitly in the following sections. Furthurmore, we also check all eigenvalue are non-negative to ensure their absolute values do not exceed the upper bounds.

\section{The verification of upper bounds on spectral radii} 
In this section, we will check each eigenvalue of the blocks is smaller than the upper bounds.

\subsection{The identity block} \label{sec: The identity block spectral radius}
The identity block has $9$ different eigenvalues as
\begin{flalign} \label{eqn: The eigenvalues of identity block}
    &\left\{\frac{\left(B^5-5 B^3+4 B\right) d \left(B^2 d^2-9\right)}{B \left(B^2
   d^2-3\right) \left(B^2 d^2-2\right) \left(B^2 d^2-1\right)},\frac{\left(B^5-5 B^3+4
   B\right) d^2 \left(B^2 d^2-9\right)}{B \left(B^2 d^2-3\right) \left(B^2 d^2-2\right)
   \left(B^2 d^2-1\right)},\right.\notag \\
   &\frac{\left(B^3-B\right) \left(B^4 d^4-13 B^2
   d^2+36\right)}{B \left(B^2 d^2-3\right) \left(B^2 d^2-2\right) \left(B^2
   d^2-1\right)},\frac{\left(B^2-1\right) d \left(B^4 d^4-13 B^2d^2+36\right)}{\left(B^2 d^2-3\right) \left(B^2 d^2-2\right) \left(B^2
   d^2-1\right)},\frac{B^4 d^4-13 B^2 d^2+36}{\left(B^2 d^2-3\right) \left(B^2
   d^2-2\right)},\notag\\
   &\frac{A_{1}-\left(B^3-B\right) \left(B^2 d^3-B^2 d^2-9 d+9\right) \sqrt{A_{2}}}{A_{3}},\frac{A_{1}+\left(B^3-B\right) \left(B^2 d^3-B^2 d^2-9 d+9\right)
   \sqrt{A_{2}}}{A_{3}}\notag\\
   &\left. \frac{A_{4}-\left(B^3-B\right) (d-1) d \sqrt{A_{5}}}{A_{6}},\frac{A_{4}+\left(B^3-B\right) (d-1) d \sqrt{A_{5}}}{A_{6}}\right\}\;,
\end{flalign}
where the polynomials $A_{1}\sim A_{6}$ are
\begin{flalign}
    A_{1} = &B^7 \left(d^5+d^4\right)+B^5 \left(-3 d^5-3 d^4-11 d^3-11 d^2\right)+B^3 \left(2 d^5+2
   d^4+29 d^3+29 d^2+18 d+18\right)\notag\\
   &+B \left(-18 d^3-18 d^2-18 d-18\right)\notag\\
    A_{2} = &B^4 d^4+B^2 \left(-4 d^4-4 d^2\right)+4 d^4+16 d^3+40 d^2+16 d+4\notag\\
    A_{3} = &2 B d \left(B^2 d^2-3\right) \left(B^2 d^2-2\right) \left(B^2 d^2-1\right) \notag\\
    A_{4} = &B^7 \left(d^5+d^4\right)+B^5 \left(-15 d^4-14 d^3+d^2\right)+B^3 \left(-d^5+44 d^4+20
   d^3+5 d^2+30 d\right) \notag\\
   &+B \left(-30 d^4-6 d^3-6 d^2-30 d\right)\notag\\
   A_{5} = &B^8 d^6+B^6 \left(2 d^6+30 d^5+2 d^4\right)+B^4 \left(d^6-110 d^5+19 d^4-110
   d^3+d^2\right)\notag\\
   &+B^2 \left(60 d^5-708 d^4-600 d^3-708 d^2+60 d\right)+900 d^4+2160
   d^3+3816 d^2+2160 d+900\notag\\
   A_{6}= &2 B d \left(B^2 d^2-3\right) \left(B^2 d^2-2\right) \left(B^2 d^2-1\right)\;.
\end{flalign}

By observation, we notice that the 2nd and 4th eigenvalues are a constant $d$ times the 1st and 3rd eigenvalues, respectively, so we only need to consider the 2nd and 4th eigenvalues because $d\geq 2$. The 5th eigenvalue is the spectral radius which satisfies the bound as we prove in the main text. The 6th, 7th eigenvalues and 8th, 9th eigenvalues are paris of solutions of quadratic equations, so we only need to check the larger ones, which are the 7th and 9th eigenvalues (this will be verified later). Now the problem is simplified to verifying that the 2nd, 4th, 7th and 9th eigenvalues are bounded by $1$.

\subsubsection{The 2nd eigenvalue}
To prove the 2nd eigenvalue is bounded by $1$ is equivalent to proving (see the supplementary Mathematica notebook)
\begin{flalign}
    B^7 \left(d^6-d^4\right)+B^5 \left(9 d^2-d^4\right)+B^3 \left(-4 d^4-34 d^2\right)+B
   \left(36 d^2-6\right)\geq 0\;.
\end{flalign}
By scaling method ($d\geq 2,B\geq 2$), the LHS is bounded as
\begin{flalign}
    {\rm LHS} &\geq 3B^7d^4-B^5d^4+9B^5d^2-4B^3d^4-34B^3d^2\notag\\
    &\geq 12B^5d^4-4B^3d^4+2B^5d^2\notag\\
    &\geq 44B^3+2B^5d^2 \geq 0\;.
\end{flalign}
Therefore, the 1st and 2nd eigenvalues of the identity block satisfy the upper bound.

\subsubsection{The 4th eigenvalue}
We need to prove (see the supplementary Mathematica notebook)
\begin{flalign}
    B^6 \left(d^6-d^5\right)+B^4 \left(d^5-6 d^4+13 d^3\right)+B^2 \left(-13 d^3+11 d^2-36
   d\right)+36 d-6\geq 0\;,
\end{flalign}
where the LHS is bounded as ($d\geq 2, B\geq 2$)
\begin{flalign}
    {\rm LHS} &\geq B^6d^5-2B^4d^5+13B^4d^3-13B^2d^3-14B^2d \notag\\
    & \geq 2B^4d^5+39B^2d^3-14B^2d\notag\\
    & \geq 2B^4d^5+142B^2d \geq 0\;.
\end{flalign}
So the 3rd and 4th eigenvalues satisfy the bound.

\subsubsection{The 7th eigenvalue}
We need to prove
\begin{flalign}
    &B^{12} \left(4 d^{14}-4 d^{12}-4 d^{11}+4 d^9\right)+B^{10} \left(-36 d^{12}+12
   d^{11}+68 d^{10}+44 d^9-88 d^7\right)\notag \\
   &+B^8 \left(-8 d^{12}-8 d^{11}+44 d^{10}-152
   d^9-380 d^8+148 d^7+612 d^5\right)\notag\\
   &+B^6 \left(120 d^{10}+104 d^9+324 d^8+108 d^7+940
   d^6-2732 d^5-1296 d^3\right)\notag\\&+B^4 \left(-520 d^8-168 d^7-1008 d^6+3728 d^5-1056
   d^4+6720 d^3\right)\notag\\
   &+B^2 \left(840 d^6-1608 d^5+960 d^4-10176 d^3+432 d^2+432
   d\right)-432 d^4+4752 d^3-288 d^2-432 d\geq 0\;.
\end{flalign}
The LHS is bounded as ($d\geq 2, B\geq 2$)
\begin{flalign}
    {\rm LHS} &\geq 4B^{12}d^{14}-6B^{12}d^{12}-39B^{10}d^{12}+4B^{12}d^{9}\notag\\
    &\geq \frac{5}{2}B^{12}d^{14}-39B^{10}d^{12} \geq 121B^{10}d^{12} \geq 0\;.
\end{flalign}
The 6th and 7th eigenvalues satisfy the bound. For the same reason in the main text, when we square both sides of the inequality, we should check $A_{1}\geq 0$, $A_{2}\geq0$, $A_{3}-A_{1}\geq 0$ and $B^2d^3-B^2d^2-9d+9 \geq 0$.

\textcircled{1} To prove $A_{1}\geq 0$, we need to prove
\begin{flalign}
    &B^7 \left(d^5+d^4\right)+B^5 \left(-3 d^5-3 d^4-11 d^3-11 d^2\right)+B^3 \left(2 d^5+2
   d^4+29 d^3+29 d^2+18 d+18\right)\notag\\
   &+B \left(-18 d^3-18 d^2-18 d-18\right)\geq 0\;.
\end{flalign}

\textbf{a.} When $d\geq 2, B\geq 3$, we have
\begin{flalign}
    A_{1}\geq B^{5}(6d^5+6d^4-11d^3-11d^2) \geq 13B^5(d^3+d^2)\geq 0
\end{flalign}

\textbf{b.} When $d\geq 2, B= 2$, we have
\begin{flalign}
    A_{1} = 48d^5+48d^4-156d^3-156d^2+108d+108 \geq 36(d^3+d^2)\geq 0\;,
\end{flalign}
so we have proved $A_{1}\geq 0$.

\textcircled{2} To prove $A_{2}\geq 0$, we need to prove
\begin{flalign}
    B^4 d^4+B^2 \left(-4 d^4-4 d^2\right)+4 d^4+16 d^3+40 d^2+16 d+4 \geq 0\;.
\end{flalign}

\textbf{a.} When $d\geq 2, B\geq 3$, we have
\begin{flalign}
    A_{2} \geq 5B^2d^4-4B^2d^2 \geq 16B^2d^2 \geq 0\;.
\end{flalign}

\textbf{b.} When $d\geq 2, B=2$, we have
\begin{flalign}
    A_{2} = 4d^4+16d^3+24d^2+16d+4 \geq 0\;,
\end{flalign}
so $A_{2} \geq 0$.

\textcircled{3} To prove $A_{3}-A_{1}\geq 0$, we need to prove
\begin{flalign}
    &B^7 \left(2 d^7-d^5-d^4\right)+B^5 \left(-9 d^5+3 d^4+11 d^3+11 d^2\right)\notag\\
    &+B^3 \left(-2
   d^5-2 d^4-7 d^3-29 d^2-18 d-18\right)+B \left(18 d^3+18 d^2+6 d+18\right) \geq 0;.
\end{flalign}

When $d\geq 2, B\geq 2$, we have
\begin{flalign}
    A_{3}-A_{1} &\geq \frac{5}{2}B^7d^5-9B^5d^5+3B^5d^4+B^3(-2d^4-2d^4-7d^3-29d^2-18d-18)\notag\\
    &\geq B^5d^5-2B^3d^5+10B^3d^4-7B^3d^3-29B6=^3d^2-18B^3d-18B^3 \notag\\
    &\geq 45B^3d^2-18B^3d-18B^3 \geq 126B^3 \geq 0\;,
\end{flalign}
which completes the proof.

\textcircled{4} For $B^2d^3-B^2d^2-9d+9$

\textbf{a.} When $d\geq 2, B\geq 3$, we have
\begin{flalign}
    B^2d^3-B^2d^2-9d+9 \geq 9d+9 \geq 0\;.
\end{flalign}

\textbf{b.} When $d\geq 3, B = 2$, we have
\begin{flalign}
    B^2d^3-B^2d^2-9d+9 = 4d^3-4d^2-9d+9 \geq 15d+9 \geq 0\;.
\end{flalign}

\textbf{c.} When $d=2, B=2$, we have
\begin{flalign}
    B^2d^3-B^2d^2-9d+9 = 7 \geq 0\;,
\end{flalign}
so $B^2d^3-B^2d^2-9d+9 \geq 0$.
\subsubsection{The 9th eigenvalue}

We need to prove
\begin{flalign}
    &B^{12} \left(4 d^{14}-4 d^{12}-4 d^{11}+4 d^9\right)+B^{10} \left(-48 d^{12}+60
   d^{11}+80 d^{10}-40 d^9-52 d^7\right)\notag\\
   &+B^8 \left(4 d^{12}-176 d^{11}+152 d^{10}-128
   d^9-500 d^8+760 d^7+144 d^5\right)\notag\\
   &+B^6 \left(120 d^{11}+740 d^9+24 d^8-2496 d^7+1360
   d^6-2180 d^5\right)\notag\\
   &+B^4 \left(-576 d^9-100 d^8+2340 d^7-828 d^6+8492 d^5-1656 d^4+24
   d^3\right)\notag\\
   &+B^2 \left(-552 d^7+240 d^6-10920 d^5+1272 d^4+120 d^3+720 d^2\right)+4464
   d^5-144 d^4-144 d^3-576 d^2 \geq 0\;.
\end{flalign}
The LHS is bounded as ($d\geq 2, B\geq 2$)
\begin{flalign}
    {\rm LHS} &\geq 104B^8d^{11}-576B^4d^9 - 100B^4d^8 -552B^2d^7 + 240 B^2d^6 -10920 B^2d^5 \notag\\
    &\geq 11784B^4d^8+240B^2d^6 - 10920 B^2 d^5 \geq 183324B^2d^6 \geq 0\;.
\end{flalign}
The 8th and 9th eigenvalues satisfy the bound. We also need to prove $A_{4}\geq 0, A_{5}\geq 0, A_{6}-A_{4}\geq 0$.

\textcircled{1} For $A_{4}\geq 0$

\textbf{a.} When $d\geq 2, B\geq 3$, we have
\begin{flalign}
    A_{4} &\geq 9B^5d^5-6B^5d^4-14B^5d^3-B^3d^5 \notag\\
    &\geq 6B^5d^3+B^5d^5-B^3d^5 \geq 6B^5d^3+8B^3d^5 \geq 0\;.
\end{flalign}

\textbf{b.} When $d\geq 2, B=2$, we have
\begin{flalign}
    A_{4}=120 d^5-60 d^4-300 d^3+60 d^2+180 d \geq 180d^4-300d^3\geq 60d^3 \geq 0\;,
\end{flalign}
so $A_{4}\geq 0$.

\textcircled{2} For $A_{5} \geq 0$

\textbf{a.} When $d\geq 2, B\geq 3$, we have
\begin{flalign}
    A_{5} &\geq 100B^4d^6+160B^4d^5+2B^6d^4-36B^4d^4-588B^2d^4-600B^2d^3-708B^2d^2 \notag\\
    &\geq 402B^4d^4-588B^2d^4-600B^2d^3-708B^2d^2\notag\\
    &\geq 3030B^2d^4-600B^2d^3-708B^2d^2 \geq 4692B^2d^2 \geq 0
\end{flalign}

\textbf{b.} When $d\geq 3, B =2 $, we have
\begin{flalign}
    A_{5} = 400 d^6+400 d^5-1500 d^4-2000 d^3+1000 d^2+2400 d+900 \geq 1900d^4+1600d^3 \geq 0\;.
\end{flalign}

\textbf{c.} When $d=2, B=2$, we have
\begin{flalign}
    A_{5} = 8100 \geq 0\;,
\end{flalign}
so $A_{5} \geq 0$.

\textcircled{3} For $A_6-A_4 \geq 0$, when $d\geq 2, B\geq 2$, we have
\begin{flalign}
    A_{6}-A_{4}&\geq B^{7}(2d^7-\frac{3}{2}d^5)+B^5(-12d^5+15d^4) +B^3(-42d^4-16d^2)\notag\\
    &\geq \frac{13}{2}B^7d^5 - 12B^5d^5+18B^3d^4 - 16B^3d^2 \geq 14B^5d^5+56B^3d^2 \geq 0\;,
\end{flalign}
so $A_{6}-A_{4}\geq 0$.
So far, we have proved all eigenvalues of the identity block are bounded by $1$.

\subsection{The $O_{1}$ block}
The $O_{1}$ block has 5 distinct eigenvalues as
\begin{flalign}
    &\left\{0,\frac{\left(B^3-B\right) \left(d^3 B^4-4 d^2 B^2-9 d B^2+36\right)}{B
   \left(B^2 d^2-3\right) \left(B^2 d^2-2\right) \left(B^2 d^2-1\right)},\frac{C_{1}-\left(B^3-B\right) \left(B^2 d^3-B^2 d^2-9
   d+9\right) \sqrt{C_{2}}}{C_{3}},\right.\notag\\
   &\left. \frac{C_{1}+\left(B^3-B\right) \left(B^2 d^3-B^2 d^2-9 d+9\right) \sqrt{C_{2}}}{C_{3}},\frac{C_{4}-B(d-1) d \sqrt{C_{5}}
   }{C_{6}},\frac{C_{4}+B(d-1) d \sqrt{C_{5}} }{C_{6}}\right\}\;,
\end{flalign}
where the polynomials $C_{1}\sim C_{6}$ are
\begin{flalign}
    C_{1} = &B^7 \left(d^5+d^4\right)+B^5 \left(-d^5-5 d^4-11 d^3-11 d^2\right)+B^3 \left(4 d^4+11
   d^3+47 d^2+18 d+18\right)\notag\\
   &+B \left(-36 d^2-18 d-18\right)\notag\\
   C_{2} = &B^4 d^4-4 B^2 d^2+16 d^2+16 d+4\notag\\
   C_{3} = &2 B d \left(B^2 d^2-3\right) \left(B^2 d^2-2\right) \left(B^2 d^2-1\right)\notag\\
   C_{4} = &B^7 \left(d^5+d^4\right)+B^5 \left(2 d^5-5 d^4-26 d^3+d^2\right)+B^3 \left(-8 d^4+23
   d^3+65 d^2+18 d\right)\notag\\
   &+B \left(-18 d^2-54 d\right)\notag\\
   C_{5} = &B^{12} d^6+B^{10} \left(4 d^6+6 d^5+2 d^4\right)+B^8 \left(4 d^6-36 d^5-213 d^4-158
   d^3+d^2\right)\notag\\&+B^6 \left(-24 d^5+156 d^4+1078 d^3+1282 d^2+36 d\right)+B^4 \left(64
   d^4-488 d^3-2075 d^2-2736 d+324\right)\notag\\&
   +B^2 \left(288 d^2+1836 d+648\right)+324\notag\\
   C_{6} = &2 B d \left(B^2 d^2-3\right) \left(B^2 d^2-2\right) \left(B^2 d^2-1\right)\;.
\end{flalign}
By observation and the same reason in the previous section, we only need to check the 2nd and 4th eigenvalues are bounded by $2/d^2$.

\subsubsection{The 2nd eigenvalue}

If the 2nd eigenvalue is bounded by $2/d^2$, we have
\begin{flalign}
    B^6 \left(2 d^6-d^5\right)+B^4 \left(d^5-8 d^4+9 d^3\right)+B^2 \left(-4 d^4-9 d^3-14
   d^2\right)+36 d^2-12 \geq 0\;,
\end{flalign}
where the LHS of the inequality is bounded as ($d\geq 2,B\geq 2$)
\begin{flalign}
    {\rm LHS} \geq 18B^4d^4-4B^2d^4-9B^2d^3-14B^2d^2 \geq 96B^2d^2\geq 0\;.
\end{flalign}
So the 2nd eigenvalue satisfies the bound.

\subsubsection{The 4th eigenvalue}
We need to prove
\begin{flalign}
    &B^{12} \left(8 d^{14}-4 d^{13}\right)+B^{10} \left(8 d^{14}+24 d^{13}-64 d^{12}+48
   d^{11}\right)+B^8 \left(-12 d^{13}-120 d^{12}-264 d^{11}+344 d^{10}-148
   d^9\right)\notag\\
   &+B^6 \left(-8 d^{13}-8 d^{12}+40 d^{11}+408 d^{10}+392 d^9-1648 d^8+584
   d^7\right)\notag\\
   &+B^4 \left(176 d^{11}+176 d^{10}+980 d^9+568 d^8-56 d^7+3568 d^6-2112
   d^5\right)\notag\\
   &+B^2 \left(-1224 d^9-1224 d^8-3120 d^7-3072 d^6+3840 d^5-1248 d^4+864
   d^3\right)\notag\\
   &+2592 d^7+2592 d^6-1728 d^5-864 d^4-864 d^3+576 d^2 \geq 0\;,
\end{flalign}
where the LHS of the inequality is bounded as ($d\geq 2, B\geq 2$)
\begin{flalign}
    {\rm LHS} &\geq 56B^{10}d^{13}-138B^8d^{13}-12B^{6}d^{13}-3096B^2d^{9}\notag\\
    &\geq 332B^6d^{13}-3096B^2d^9 \geq 81896B^2d^9 \geq 0\;.
\end{flalign}
So the 3rd and 4th eigenvalues satisfy the bound. Then we should check $C_{1}\geq 0, C_{2}\geq 0$ and $2C_{3}-d^2C_{1}\geq 0$.

\textcircled{1} For $C_{1}\geq 0$

\textbf{a.} When $d\geq 2, B\geq 3$, we have
\begin{flalign}
    C_{1} \geq 8B^5d^5+4B^3d^4-11B^5d^3-11B^5d^2 \geq 21B^5d^3+5B^5d^2 \geq 0\;.
\end{flalign}

\textbf{b.} When $d\geq 2,B=2$, we have
\begin{flalign}
    C_{1} = 96d^5-264d^3-48d^2+108d+108 \geq 120d^3-48d^2 \geq 192d^2 \geq 0\;,
\end{flalign}
so $C_{1}\geq 0$.

\textcircled{2} For $C_{2}\geq 0$, we have ($d\geq 2, B\geq 2$)
\begin{flalign}
    C_{2}\geq 12B^2d^2 \geq 0\;.
\end{flalign}

\textcircled{3} For $2C_{3}-d^2 C_{1}\geq 0$, we have ($d\geq 2, B\geq 2$)
\begin{flalign}
    2C_{3}-d^2 C_{1}&\geq 5B^7d^6+B^5d^5-11B^5d^4-4B^3d^6-11B^3d^5 - 47B^3d^4\notag\\
    &\geq B^5d^5 + 13B^3d^5 + 165B^3d^4 \geq 0\;.
\end{flalign}
So far, we have proved all eigenvalues of the $O_{1}$ block is bounded by $2/d^2$.

\subsection{The $O_{2}$ block}
The distinct eigenvalues of the $O_{2}$ block are 
\begin{flalign}
    &\left\{0,\frac{\left(B^3-B\right) \left(B^4 d^3-4 B^2 d^2-9 B^2 d+36\right)}{B
   \left(B^2 d^2-3\right) \left(B^2 d^2-2\right) \left(B^2 d^2-1\right)},\frac{B^6
   d^3+5 B^4 d^3-20 B^4 d^2+B^4 d-16 B^2 d^2+65 B^2 d-36}{\left(B^2 d^2-3\right)
   \left(B^2 d^2-2\right) \left(B^2 d^2-1\right)},\right.\notag\\
   &\left.\frac{\left(B^3-B\right) \left(B^4
   d^4-2 B^2 d^3-11 B^2 d^2+18 d+18\right)}{B d \left(B^2 d^2-3\right) \left(B^2
   d^2-2\right) \left(B^2 d^2-1\right)}\right\}\;,
\end{flalign}
where the 3rd eigenvalue is the spectral radius, so we only need to check the 2nd and 4th eigenvalues are bounded by $3/d^3$.

\subsubsection{The 2nd eigenvalue}
If the 2nd eigenvalue is bounded by $3/d^3$, we have
\begin{flalign}
    2 B^6 d^6+B^4 \left(d^6+4 d^5-9 d^4\right)+B^2 \left(-4 d^5-9 d^4-36 d^3+33
   d^2\right)+36 d^3-18\geq 0\;,
\end{flalign}
where the LHS of the inequality is bounded as ($d\geq 2,B\geq 2$)
\begin{flalign}
    {\rm LHS} &\geq 7B^4d^4 + 28B^2d^5 - 9B^2d^4 - 36d^3 \notag\\
    &\geq 7B^4d^4 + 47 B^2d^4 - 36d^3 \geq 7B^4d^4 + 340d^3 \geq 0\;.
\end{flalign}
So the 2nd eigenvalue is bounded by $3/d^3$.

\subsubsection{The 4th eigenvalue}
We need to prove
\begin{flalign}
    2 B^6 d^6+B^4 \left(d^6+2 d^5-7 d^4\right)+B^2 \left(-2 d^5-11 d^4-18 d^3+15
   d^2\right)+18 d^3+18 d^2-18 \geq 0\;,
\end{flalign}
where the LHS of the inequality satiefies
\begin{flalign}
    {\rm LHS}&\geq 2B^6d^6 + B^4d^6 + 2B^4d^5 - 7B^4d^4 -2B^2d^5-20B^2d^4 \notag\\
    &\geq 9B^4d^6-3B^4d^4-2B^2d^5 - 20B^2d^4 \notag\\
    &\geq B^4d^4+62B^2d^5-20B^2d^4 \geq B^4d^4 + 124B^2d^4 \geq 0 \;.
\end{flalign}
So the 4th eigenvalue is bounded by $3/d^3$. 

So far, we have proved all eigenvalues of the interaction blocks are bounded as expected.

\subsection{The verification of upper bound on $O_{1}$ block when $d=2$}
When $d=2$, the $O_{1}$ block has a better bound $1/d^2$ on its spectral radius, so we verify it in this section.

\subsubsection{The 2nd eigenvalue}
To prove the 2nd eigenvalue is bounded by $1/d^2$, we need to prove
\begin{flalign}
    32 B^6+72 B^4-236 B^2+138 \geq 0\;,
\end{flalign}
where the LHS of the inequality satisfies ($B\geq 2$)
\begin{flalign}
    {\rm LHS} \geq 32B^6+52B^2 \geq 0\;,
\end{flalign}
which completes the proof.

\subsubsection{The 4th eigenvalue}
We need to prove
\begin{flalign}
    106496 B^{10}+57344 B^8-1271296 B^6+2403072 B^4-1755264 B^2+460224\geq 0\;,
\end{flalign}
where the LHS of the inequality satisfies
\begin{flalign}
    {\rm LHS} \geq 662016B^6 \geq 0\;,
\end{flalign}
which completes the proof.

\subsubsection{The 6th eigenvalue}
We need to prove
\begin{flalign}
    8192 B^{10}+111104 B^8-532096 B^6+813312 B^4-516288 B^2+115920\geq 0\;,
\end{flalign}
where the LHS satisfies
\begin{flalign}
    {\rm LHS} \geq 43392B^6 \geq 0\;,
\end{flalign}
which completes the proof. So far we have proved when $d=2$, the spectral radius of $O_1$ block is bounded by $1/d^2$.

\section{The verification of non-negativity of all eigenvalues}
In this section, we will check each eigenvalue of the blocks is non-negative.

\subsection{The identity block}
We only need to check the 2nd, 4th, 5th, 6th and 8th eigenvalues of the identity block by the same reason in Section. \ref{sec: The identity block spectral radius}. we only need to check the numerators because the common denominator is positive.

\subsubsection{The 2nd eigenvalue}
To prove the non-negativity of the 2nd eigenvalue, we need to prove
\begin{flalign}
    B^7 d^4+B^5 \left(-5 d^4-9 d^2\right)+B^3 \left(4 d^4+45 d^2\right)-36 B d^2 \geq 0\;,
\end{flalign}
where the LHS satisfies ($d\geq 2, B\geq 2$)
\begin{flalign}
    {\rm LHS} \geq 4B^5d^4-9B^5d^2 \geq 7B^5d^2 \geq 0\;,
\end{flalign}
which completes the proof.

\subsubsection{The 4th eigenvalue}
We need to prove
\begin{flalign} \label{eqn: 4th eigenvalue non-negativity}
    B^4 d^4-13 B^2 d^2+36 \geq 0
\end{flalign}
which is obvious when $d\geq 2, B\geq 2$, so complete the proof.

\subsubsection{The 5th eigenvalue}
The requirement is the same as \eqref{eqn: 4th eigenvalue non-negativity}, so we complete the proof.

\subsubsection{The 6th eigenvalue}
We need to prove $A_{1}^2 - \left((B^3-B)(B^2d^3-B^2d^2-9d+9)\sqrt{A_{2}}\right)^2 \geq 0$\;, or equivalently to
\begin{flalign}
&4 B^{12} d^9+B^{10} \left(-24 d^9-88 d^7\right)+B^8 \left(36 d^9+528 d^7+612
   d^5\right)+\notag\\
   &B^6 \left(-16 d^9-792 d^7-3672 d^5-1296 d^3\right)+B^4 \left(352 d^7+5508
   d^5+7776 d^3\right)\notag\\
   &+B^2 \left(-2448 d^5-11664 d^3\right)+5184 d^3 \notag \geq 0 \;.
\end{flalign}

\textbf{a.} When $d\geq 2, B\geq 4$, the LHS satisfies
\begin{flalign}
    {\rm LHS} \geq 40B^10d^{9} - 88B^{10}d^{7}\geq 72B^{10}d^{7} \geq 0\;.
\end{flalign}

\textbf{b.} When $d\geq 2, B=3$, we have
\begin{flalign}
    {\rm LHS} = 933120 d^9-2280960 d^7+1762560 d^5-414720 d^3 \geq 1451520 d^7 \geq 0\;.
\end{flalign}

\textbf{c.} When $d\geq 2, B =2$, we have
\begin{flalign}
    {\rm LHS} = 0 \geq 0\;.
\end{flalign}
So the 6th and 7th eigenvalues are non-negative.

\subsubsection{The 8th eigenvalue}
We need to prove 
\begin{flalign}
    &4 B^{12} d^9+B^{10} \left(-60 d^9-52 d^7\right)+B^8 \left(252 d^9+780 d^7+144
   d^5\right)\notag\\
   &+B^6 \left(-340 d^9-3276 d^7-2160 d^5\right)+B^4 \left(144 d^9+4420
   d^7+9072 d^5\right)\notag\\
   &+B^2 \left(-1872 d^7-12240 d^5\right)+5184 d^5 \geq 0 \;.
\end{flalign}
We can easily check the polynomial after $B^{8}$ term is positive when $d\geq 2, B\geq 2$, so we only need to focus on the $B^{12}$ and $B^{10}$ terms.

\textbf{a.} When $d\geq 2, B\geq 5$, we have
\begin{flalign}
    {\rm LHS} \geq 40B^{10}d^{9}-52B^{10}d^{7} \geq 108B^{10}d^{7}\geq 0
\end{flalign}

\textbf{b.} When $d \geq 2, B = 4$, we have
\begin{flalign}
    {\rm LHS} = 19353600d^9-15724800d^7 + 2721600 d^5 \geq 61689600 d^7 \geq 0\;.
\end{flalign}

\textbf{c.} When $d\geq 2, B = 3$, we have
\begin{flalign}
    {\rm LHS} = 0 \geq 0\;.
\end{flalign}

\textbf{d.} When $d\geq 2, B =2$, we have
\begin{flalign}
    {\rm LHS} = 0 \geq 0\;.
\end{flalign}
So the 8th and 9th eigenvalues are non-negative.

\subsection{The $O_{1}$ block}
We check the 2nd, 3rd and 5th eigenvalues are non-negative.

\subsubsection{The 2nd eigenvalue}
We need to prove
\begin{flalign}
    \left(B^2-1\right) (B d-3) (B d+3) \left(B^2 d-4\right) \geq 0 \;,
\end{flalign}
which is obvious since each parenthesis is positive, so we completes the proof.

\subsubsection{The 3rd eigenvalue}
We need to prove
\begin{flalign}
    &4 B^{12} d^9+B^{10} \left(-16 d^9-8 d^8-88 d^7\right)+B^8 \left(20 d^9+16 d^8+352
   d^7+176 d^6+612 d^5\right)\notag\\
   &+B^6 \left(-8 d^9-8 d^8-440 d^7-352 d^6-2448 d^5-1224
   d^4-1296 d^3\right)\notag\\
   &+B^4 \left(176 d^7+176 d^6+3060 d^5+2448 d^4+5184 d^3+2592
   d^2\right)\notag\\
   &+B^2 \left(-1224 d^5-1224 d^4-6480 d^3-5184 d^2\right)+2592 d^3+2592 d^2 \geq 0\;,
\end{flalign}
whose terms with lower order than $B^{10}$ are always positive when $d\geq 2, B\geq 2$.

\textbf{a.} When $d\geq 2, B\geq 4$, we have
\begin{flalign}
    {\rm LHS} \geq 48B^{10}d^{9} - 8B^{10}d^{8}-88B^{10}d^{7} \geq 88B^{10}d^{7}\geq 0\;.
\end{flalign}

\textbf{b.} When $d\geq 2, B = 3$, we have
\begin{flalign}
    {\rm LHS} \geq 1306368d^9 - 373248d^8-3193344d^7 \geq 1285632d^7 \geq 0\;.
\end{flalign}

\textbf{c.} When $d\geq 3, B = 2$, we have
\begin{flalign}
    {\rm LHS} \geq 4608d^9-4608d^8-25344d^7 \geq 9216d^8-25344d^7 \geq 2304d^7 \geq 0 \;.
\end{flalign}

\textbf{d.} When $d=2, B=2$, we have
\begin{flalign}
    {\rm LHS} = 169344 \geq 0\;.
\end{flalign}
So we have proved the 3rd and 4th eigenvalues are non-negative.

\subsubsection{The 5th eigenvalue}
We need to prove
\begin{flalign}
    &4 B^{12} d^7+B^{10} \left(-4 d^7-56 d^6-52 d^5\right)+B^8 \left(8 d^7+88 d^6+208
   d^5+728 d^4+144 d^3\right)\notag\\
   &+B^6 \left(-8 d^7-32 d^6-332 d^5-1216 d^4-2172 d^3-2016
   d^2\right)+B^4 \left(176 d^5+488 d^4+3252 d^3+4104 d^2+5616 d\right)\notag\\
   &+B^2 \left(-1224
   d^3-2088 d^2-8208 d-2592\right)+2592 d+2592\geq 0 \;,
\end{flalign}
where the sum of terms with order lower than $B^{10}$ is positive when $d\geq 2, B\geq 2$.

\textbf{a.} When $d\geq 2,B\geq 4$, we have
\begin{flalign}
    {\rm LHS} \geq 60B^{10}d^{7} - 56B^{10}d^{6}-52B^{10}d^{5} \geq 76B^{10}d^5\geq 0\;.
\end{flalign}

\textbf{b.} When $d\geq 2, B=3$, we have
\begin{flalign}
    {\rm LHS} &= 1936224 d^7-2752704 d^6-1933632 d^5+3929472 d^4-386208 d^3-1156032 d^2+383616 d-20736 \notag\\ &\geq 305856d^5 \geq 0 \;.
\end{flalign}

\textbf{c.} When $d\geq 4, B=2$, we have
\begin{flalign}
    {\rm LHS} & = 13824 d^7-36864 d^6-18432 d^5+116352 d^4-55008 d^3-71712 d^2+59616 d-7776 \geq 55296d^5 \geq 0 \;.
\end{flalign}

\textbf{d.} When $d=3,B=2$, we have
\begin{flalign}
    {\rm LHS} = 6345216 \geq 0\;.
\end{flalign}

\textbf{e.} When $d=2,B=2$, we have
\begin{flalign}
    {\rm LHS} = 66528 \geq 0\;.
\end{flalign}

So the 5th and 6th eigenvalues are non-negative. So far, all eigenvalues of the $O_1$ block are non-negative.

\subsection{The $O_{2}$ block}
We check the 2nd, 3rd and 4th eigenvalues are non-negative.

\subsubsection{The 2nd eigenvalue}
We need to prove
\begin{flalign}
    B^4 d^3-4 B^2 d^2-9 B^2 d+36 \geq 0\;.
\end{flalign}

\textbf{a.} When $d\geq 2, B\geq 3$, we have
\begin{flalign}
    {\rm LHS} \geq 14B^2d^2-9B^2d \geq 19B^2d \geq 0 \;.
\end{flalign}

\textbf{b.} When $d\geq 3, B=2$, we have
\begin{flalign}
    {\rm LHS} = 16 d^3-16 d^2-36 d+36 \geq 60d \geq 0\;.
\end{flalign}

\textbf{c.} When $d=2, B=2$, we have
\begin{flalign}
    {\rm LHS} = 28 \geq 0\;.
\end{flalign}
So the 2nd eigenvalue is non-negative.

\subsubsection{The 3rd eigenvalue}
We need to prove
\begin{flalign}
    B^6 d^3+B^4 \left(5 d^3-20 d^2+d\right)+B^2 \left(65 d-16 d^2\right)-36 \geq 0\;.
\end{flalign}

\textbf{a.} When $d\geq 2, B\geq 3$, we have
\begin{flalign}
    {\rm LHS} \geq 14B^4d^3-20B^4d^2 - 16B^2d^2 \geq 56B^2d^2\geq 0\;.
\end{flalign}

\textbf{b.} When $d\geq 3, B=2$, we have
\begin{flalign}
    {\rm LHS} = 144 d^3-384 d^2+276 d-36 \geq 48d^2 \geq 0 \;.
\end{flalign}

\textbf{c.} When $d=2, B=2$, we have
\begin{flalign}
    {\rm LHS} = 132 \geq 0\;.
\end{flalign}
So the 3rd eigenvalue is non-negative.

\subsubsection{The 4th eigenvalue}
We need to prove 
\begin{flalign}
    B^4 d^4+B^2 \left(-2 d^3-11 d^2\right)+18 d+18 \geq 0\;.
\end{flalign}
When $d\geq 2, B\geq 2$, we have
\begin{flalign}
    {\rm LHS} \geq 6B^2d^3-11B^2d^2 \geq B^2d^2 \geq 0\;.
\end{flalign}
So the 4th eigenvalue is non-negative. So far we have proved all eigenvalues are non-negative.

\end{document}